\documentclass[fleqn,usenatbib]{mnras}
\usepackage[T1]{fontenc}
\usepackage{ae,aecompl}
\usepackage{graphicx}
\usepackage{amsmath}
\usepackage{amssymb}

\newcommand{\galaxia}{\textsc{galaxia}}
\newcommand{\kms}{\ifmmode  \,\rm km\,s^{-1} \else $\,\rm km\,s^{-1}  $ \fi }
\newcommand{\kpc}{\ifmmode  {\rm kpc}  \else ${\rm  kpc}$ \fi  }  
\newcommand{\Msun}{\ifmmode {\rm M_{\odot}} \else ${\rm M_{\odot}}\odot\odot$ \fi} 
\newcommand{\rsigma}{\sigma_r}
\newcommand{\thsigma}{\sigma_\theta}
\newcommand{\phsigma}{\sigma_\phi}

\newcommand{\vlos}{v_{\rm los}}
\newcommand{\vrot}{v_{\rm rot}}
\newcommand{\feh}{[Fe/H]}

\title[Kinematics of the inner halo of the MW]{Galactic googly\thanks{In the sport of cricket a googly is a type of delivery by a right-arm leg spin bowler that spins from off to leg in the opposite direction to a normal leg break spin.}: the rotation-metallicity bias in the inner stellar halo of the Milky Way}
\author[P. R. Kafle et al.]
{Prajwal R. Kafle,$^{1}$ \thanks{E-mail: prajwal.kafle@uwa.edu.au}
Sanjib Sharma,$^{2}$
Aaron S. G. Robotham,$^{1}$
Raj K. Pradhan, $^{3}$
\newauthor
Magda Guglielmo,$^{2}$
Luke J. M. Davies, $^{1}$
and 
Simon P. Driver,$^{1,4}$
\newauthor
\\
{\footnotesize $^{1}$ ICRAR, The University of Western Australia, 35 Stirling Highway, Crawley, WA 6009, Australia}\\
{\footnotesize $^{2}$ SIfA, A28 School of Physics, The University of Sydney, Sydney, NSW 2006, Australia}\\
{\footnotesize $^{3}$ Central Department of Physics, Tribhuvan University, Kirtipur, Kathmandu, Nepal}\\
{\footnotesize $^{4}$ SUPA, School of Physics \& Astronomy, University of St Andrews, North Haugh, St Andrews KY16 9SS, UK}\\
}

\date{Accepted 2017 June 2. Received 2017 June 2; in original form 2017 March 7}

\pubyear{2017}

\begin{document}
\label{firstpage}
\pagerange{\pageref{firstpage}--\pageref{lastpage}}
\maketitle

\begin{abstract}
The first and second moments of stellar velocities encode important information about the formation history of the Galactic halo.
However, due to the lack of tangential motion and inaccurate distances of the halo stars, the velocity moments in the Galactic halo have largely remained ``known unknowns''.
Fortunately, our off-centric position within the Galaxy allows us to estimate these moments in the galacto-centric frame using the observed radial velocities of the stars alone.
We use these velocities coupled with the Hierarchical Bayesian scheme, which allows easy marginalisation over the missing data (the proper-motion, and uncertainty-free distance and line-of-sight velocity), 
to measure the velocity dispersions, orbital anisotropy ($\beta$) and streaming motion ($\vrot$) of the halo main-sequence turn-off (MSTO) and K-giant (KG) stars in the inner stellar halo ($r\lesssim 15~\kpc$).
We study the metallicity bias in kinematics of the halo stars and observe that the
comparatively metal-rich (\feh$>-1.4$) and the metal-poor (\feh$\leq-1.4$) MSTO samples show a clear systematic difference in 
$\vrot\sim20-40\kms$, depending on how restrictive the spatial cuts to cull the disk contamination are.
The bias is also detected in KG samples but with less certainty.
Both MSTO and KG populations suggest that the inner stellar halo of the Galaxy is radially biased i.e. $\rsigma>\thsigma$ or $\phsigma$ and $\beta\simeq0.5$.
The apparent metallicity contrariety in the rotation velocity among the halo sub-populations supports the co-existence 
of multiple populations in the galactic halo that may have formed through distinct formation scenarios, i.e. in-situ versus accretion.
\end{abstract}

\begin{keywords}
Galaxy:kinematics and dynamics -- formation -- halo -- general
\end{keywords}

\section{Introduction}
The first and second moments of velocities, and the orbital anisotropy of stars, are essential parameters for the dynamical studies of astrophysical systems at all scales,
ranging from the dwarf spheroidals \citep{2009MNRAS.394L.102L,2017arXiv170505724D}, 
the Galactic halo \citep{1980MNRAS.193..295F,2001ApJ...558..666B,2004AJ....127..914S,2005MNRAS.364..433B,2009ApJ...698.1110S,2010ApJ...716....1B,2012ApJ...761...98K,2013MNRAS.430.2973K,2016MNRAS.460.1725D}, 
the Andromeda galaxy \citep{2010MNRAS.406..264W,2013ApJ...768L..33V}, distant galaxies \citep{2010MNRAS.406..264W,2012ApJ...748....2D}, 
galaxy groups and clusters \citep{1997ApJ...485L..13C,2010MNRAS.408.2442W,2015MNRAS.453.3848D} to dark mater haloes in cosmological simulations \citep{2013MNRAS.434.1576W}.  
These moments are crucial in measuring the underlying mass of the system via Jeans analysis, tracer mass formalism etc \citep{2003ApJ...583..752E,2010MNRAS.406..264W,2014ApJ...794...59K}, and 
are useful in determining the overall shape of the gravitational potential \citep{2016ApJ...816...35A,2016MNRAS.458..726L} and flattening of the dark matter halo \citep{2016MNRAS.460..329B}.
However, until recently, due to the unavailability of tangential motion and accurate distances in the case of our own Milky Way stellar halo, the moments have largely remained unknown.

With the onset of large spectroscopic endeavours such as the Sloan Extension for Galactic Understanding and Exploration \citep[SEGUE,][]{2009AJ....137.4377Y},
a sub-survey of the Sloan Digital Sky Survey \citep[SDSS,][]{2000AJ....120.1579Y} focussed on Galactic science, it is now
\footnote{For an overall summary of the earlier works such as \cite{1980MNRAS.193..295F,1989ApJ...339..126R,1994MNRAS.271...94S} etc,
we refer the reader to the review articles \cite{1993ARA&A..31..575M,2008A&ARv..15..145H,2016ARA&A..54..529B} that already cover the topic extensively.}
possible to quantify the spatio-kinematic properties of the Galactic halo beyond the solar-neighbourhood in unprecedented detail.
For example, equipped with the full phase-space coordinates of stars i.e. position, distance, line-of-sight velocity and proper-motion, 
\cite{2009MNRAS.399.1223S} study SDSS sub-dwarf stars and similarly, \cite{2010ApJ...716....1B} study SDSS main-sequence stars 
within the solar neighbourhood and find the velocity anisotropy of the local halo is radially biased ($\beta\simeq0.68$).
Note, the velocity anisotropy $\beta\in(-\infty,1)$ is a commonly used parameter to describe the orbital structure of a halo,
where $\beta<0$, $\beta=0$, and $\beta>0$ mean tangentially biased, isotropic and radially biased. 
It is given by 
\begin{equation}\label{eqn:beta}
\beta = 1 - \frac{\thsigma^2 + \phsigma^2}{2 \rsigma^2}, 
\end{equation}
where $\rsigma, \thsigma, \phsigma$ are a short-form notation for
the radial ($\sigma_{v_{r}}=\langle v_r^2 \rangle$), 
angular ($\sigma_{v_{\theta}}=\langle v_\theta^2 \rangle$) and 
azimuthal ($\sigma_{v_{\phi}}=\langle v_\phi^2 \rangle$) galacto-centric velocity dispersions in a spherical polar coordinate system.
The velocity anisotropy $\beta$ is a highly asymmetric function around the isotropy. 
Hence, it is occasionally sensible to use a modified definition of the velocity anisotropy expressed as $\beta/(2-\beta)$.
The modified velocity anisotropy is symmetric and $\in [-1,1]$, where 0 signifies an isotropic, $<0$ means a tangentially-biased, and $>0$ means a radially-biased halo.

At large distances, the proper motions of the halo stars are either unreliable or generally unavailable, which hinders a direct measurement of their velocity dispersions. 
However, our off-centric location in the Galaxy means that the galacto-centric radial ($r$) and helio-centric radial ($s$) directions are not same.
This difference is more significant in the inner halo, at a distance of $r\lesssim$ a couple of times of $R_0$, where $R_0$ is the distance of the Sun from the Galactic centre. 
Hence, in the inner halo the observed line-of-sight velocities of the stars can be expressed in terms of all three orthogonal galacto-centric velocities ($v_r, v_\theta, v_\phi$), 
or in other words the line-of-sight velocities have some contribution from the tangential galacto-centric velocities. 
Provided we have a model that well represents the distribution of the halo stars, 
we can fit a model marginalised over the unknown tangential motions, to the available 4-dimensional data (position vector and a line-of-sight velocity), and thus estimate the velocity moments of the system.
In the absence of proper-motion, the approach of estimating moments of the velocities has been extensively used to predict the kinematics of the Milky Way halo.
For example, \cite{2004AJ....127..914S,2012ApJ...761...98K,2013MNRAS.430.2973K,2014ApJ...794...59K,2015ApJ...813...89K}
fit an ellipsoidal distribution of velocities and similarly, \cite{2011MNRAS.411.1480D} 
apply an alternative power-law model to derive the halo kinematics using the marginalisation scheme.
Assuming the velocity ellipsoid, 
\cite{2012ApJ...761...98K,2013MNRAS.430.2973K} studied halo Blue Horizontal Branch stars (BHBs),
\cite{2014ApJ...794...59K} studied both BHBs and K-Giant stars (KGs), while 
\cite{2015ApJ...813...89K} analysed a mix-bag of BHB and F-type stars to cumulatively construct the velocity dispersion and anisotropy of the outer halo. 
Interestingly, \cite{2012ApJ...761...98K} found that the velocity anisotropy parameter of the 
Galactic halo is non-monotonic and has a prominent dip at a galacto-centric radius of $r\simeq18~\kpc$.
In their studies \cite{2015ApJ...813...89K} find that the value of $\beta$ at $15\lesssim R/\kpc \lesssim25$ is more tangentially biased,
which they attribute to the difference in the spatial resolutions of the data sets, and adopted marginalization technique.
A varying level of undulations in the anisotropy parameter have also been observed in simulated halos \citep{2013ApJ...773L..32R,2017arXiv170406264L}.
There are a number of proposed scenarios that could explain such a feature, e.g., 
a transition from inner to outer halo or a local shell like structure at the given radius.
Moreover, it can also be due to the un-relaxed stars dispersed from the kinematically coherent satellite galaxies that are aligned in kinematically coherent planar structures; assuming that such planar structures have strong rotation as 
suggested by \cite{2013Natur.493...62I,2015MNRAS.453.1047P,2015MNRAS.452.1052L,2015ApJ...805...67I} etc.
Recently, \cite{2017arXiv170406264L} suggest that a major merger as early as redshift $z\sim1$ can also result in a tangential dip in the value of $\beta$ over a wide range of radii.
While \cite{2015MNRAS.452.2675B} suggest that such feature in the velocity anisotropy run of the halo is a transitory phase, 
\cite{2017arXiv170406264L} conclude that such dips are long-lived in the in situ stellar halo.
In any case, there is currently no consensus view as to what causes such velocity anisotropy changes.
Finally, in the outer halo there have been recent attempts to utilise multi-epoch Hubble Space Telescope data to estimate the halo velocity dispersion.
In particular, recently \cite{2016ApJ...820...18C} use the Galactic foreground stars along the M31 galaxy and found that the halo is isotropic at $r/\kpc\sim25$.
In Figure~\ref{fig:betaprofile} we summarize the recent (this paper inclusive) measurements of the halo velocity anisotropy.

As mentioned earlier, as of yet we do not have a statistically robust sample of the halo stars with which to constrain a detailed halo kinematic map. 
This is largely due to the fact that they are difficult and inefficient to identify with current ground-based telescopes.
This makes the attempt to construct a comprehensive model for the formation history of the galactic halo a challenging task.
Observations suggest that the halo is partly formed in situ and partly by accretion, the so called dual-origin of the halo
\cite[e.g.][etc]{1992ApJS...78...87M,2007Natur.450.1020C,2010ApJ...712..692C,2011MNRAS.411.1480D,2012ApJ...746...34B,2013MNRAS.430.2973K,2017ApJ...841...59Z},
and are supported by the findings from recent cosmological-hydrodynamical simulations of galaxy formation 
\citep{2009ApJ...702.1058Z,2011MNRAS.416.2802F,2012MNRAS.420.2245M,2014MNRAS.439.3128T}.
The spatio-kinematics of the in situ component dominating the inner region of the halo 
are thought to resemble that of the disc-flattening with net prograde rotation.
The accreted component dominates the outer region of the halo, and is found to have retrograde motion.
The veracity of these claims however, have been challenged and demonstrated not to be robust \citep{2011MNRAS.415.3807S,2013MNRAS.432.2402F}.
\cite{2011MNRAS.415.3807S} identify unphysical Gaussian analysis on the azimuthal velocity distribution, 
inaccurate distances of the main-sequence stars and a lack of proper treatment of uncertainties
as the main limitations of the \cite{2007Natur.450.1020C,2010ApJ...712..692C,2012ApJ...746...34B} analysis;
for the full account of this refer to \cite{2014ApJ...786....7S}.
A re-analysis of their data also fails to find any reliable evidence for a counter-rotating halo component,
and hints that even if a distinct counter-rotating halo exists, the magnitude of the rotation must be comparatively much weaker
than the earlier claims of $40-70\kms$ \citep{2007Natur.450.1020C,2010ApJ...712..692C}. 
Furthermore, in the light of improved galactic models and increasing evidence of lighter Galaxy halo mass 
\citep[e.g.][etc]{2008ApJ...684.1143X,2014ApJ...794...59K,2014MNRAS.445.3788G,2016ApJ...829..108E} 
the impact of the model of the Galactic potential assumed in \cite{2007Natur.450.1020C,2010ApJ...712..692C} has not been explored. 
A star in a lighter Galactic halo will attain higher maximum distance above and below the Galactic plane as compared to a heavier halo. 
This will change the classification of stars into an inner and an outer halo components.
Adding to the confusion, \cite{2013MNRAS.430.2973K} find that the SEGUE BHBs do not show any radial segregation into inner prograde and outer retrograde halo.
Instead they observe a distinct dichotomy in the kinematics among the comparatively metal-rich ([Fe/H]$>-2$) and metal-poor ([Fe/H]$\leq-2$) BHB sub-samples, 
even with the new distance calibration by \cite{2013MNRAS.430.1294F}.
\cite{2013ApJ...763L..17H} report similar metallicity bias in the inner-halo kinematics whereas \cite{2013MNRAS.432.2402F} 
claim that after flagging out $\sim500$ metal-poor stars from the original BHB catalogue of \cite{2011ApJ...738...79X} 
the apparent discrepancy in the rotation signal detected in the above work diminishes. 
Moreover, \cite{2013MNRAS.432.2402F} confirm that the conclusion of a non-rotating halo holds when they utilise proper-motions of stars as well as with both a model dependent and a direct approach using de-projected line-of-sight velocities.

To further investigate the issue of halo duality, we require a large number of halo stars with reliable proper motions and robust distance estimates.
In addition it would also be informative to have halo stars from different types of stellar populations with different ages and metallicity.
In the near future, \emph{Gaia} \citep{2005MNRAS.359.1287B} will provide parallaxes and proper motions, as well as radial velocities for a large number of halo stars in the solar neighbourhood, and this will allow us to study the kinematics of the halo in greater detail. 
Similarly, chemical footprints of halo stars provided by surveys such as GALactic Archaeology with HERMES \citep[Galah,][]{2015MNRAS.449.2604D,2017MNRAS.465.3203M} will also open a completely new dimension to further test the theory.
For now, we focus on studying the kinematics of the halo K-giant stars (KGs) and halo main-sequence turn-off stars (MSTOs) using data provided by the SEGUE survey. 

The rest of the paper is organized as follows: Section~\ref{sec:data} describes our stellar halo samples. 
Section~\ref{sec:method} provides the formulary for the kinematic measurement, and tests on synthetic data are provided in Section~\ref{sec:test}.
In Section~\ref{sec:results} we present and discuss our results. 
Finally, we summarize our findings in Section~\ref{sec:conclusion}. 

\section{Data:}\label{sec:data}

\subsection{Main-sequence turn-off stars (MSTOs)}\label{sec:mstos}
\begin{figure}
   \centering
   \includegraphics[width=1.0\columnwidth]{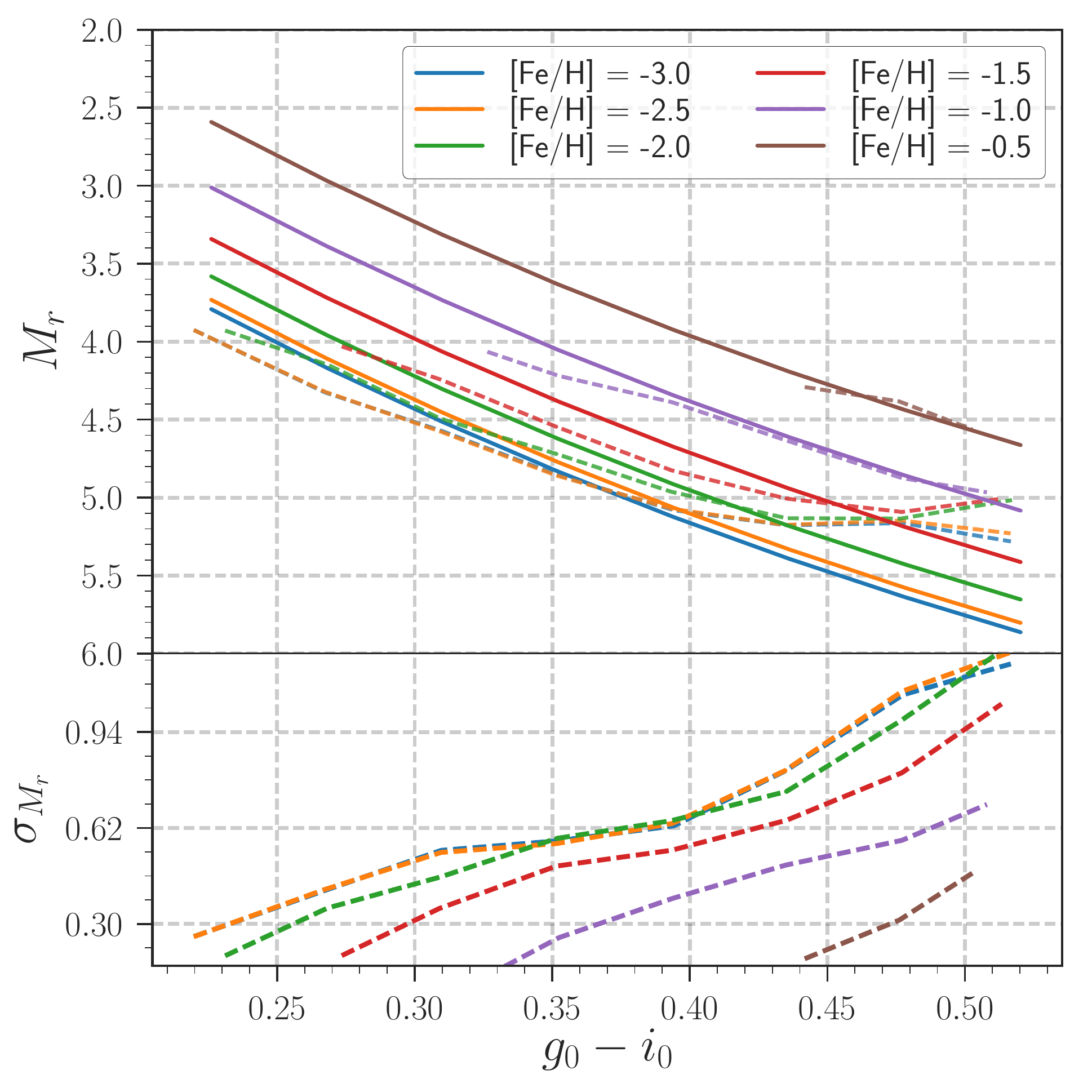}   
   \caption{Absolute magnitude-colour relation for MSTOs. 
            The solid and dashed lines in the top panel show r-band absolute magnitude, $g_0-i_0$ colour, and metallicity [Fe/H] relation 
	    obtained from \protect\cite{2008ApJ...684..287I} and \protect\citet[][\galaxia]{2011ApJ...730....3S} respectively.
	    Bottom panel shows dispersion in the r-band absolute magnitude as a function of $g_0-i_0$ colour derived 
	    using MSTOs generated from the \galaxia. 
            }
   \label{fig:Mr}
\end{figure}

\begin{figure*}
   \centering
   \includegraphics[width=1.45\columnwidth]{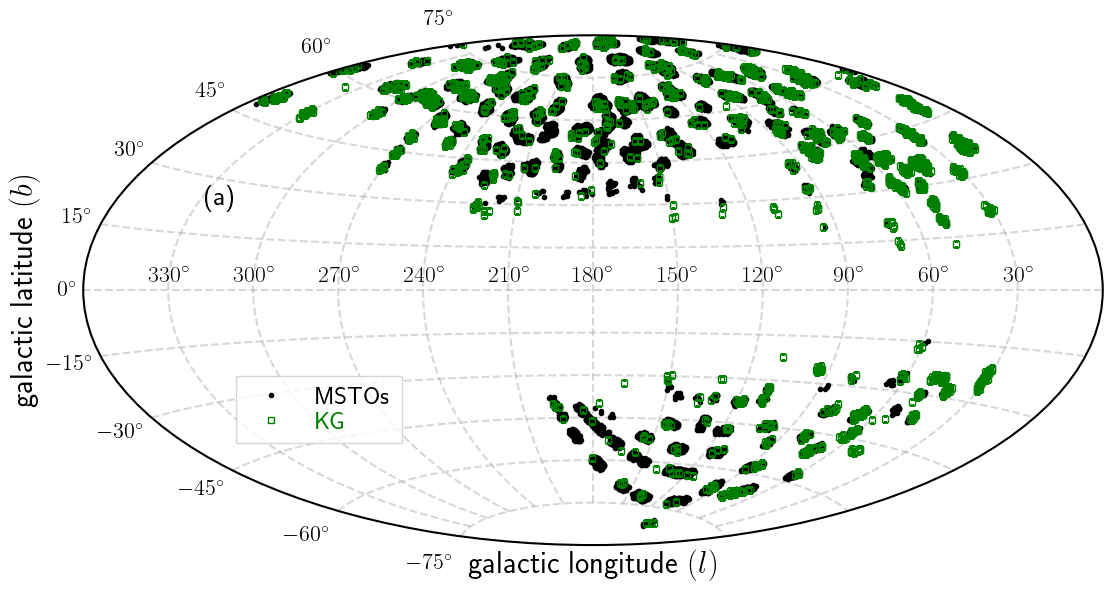}
   \includegraphics[width=0.7\columnwidth]{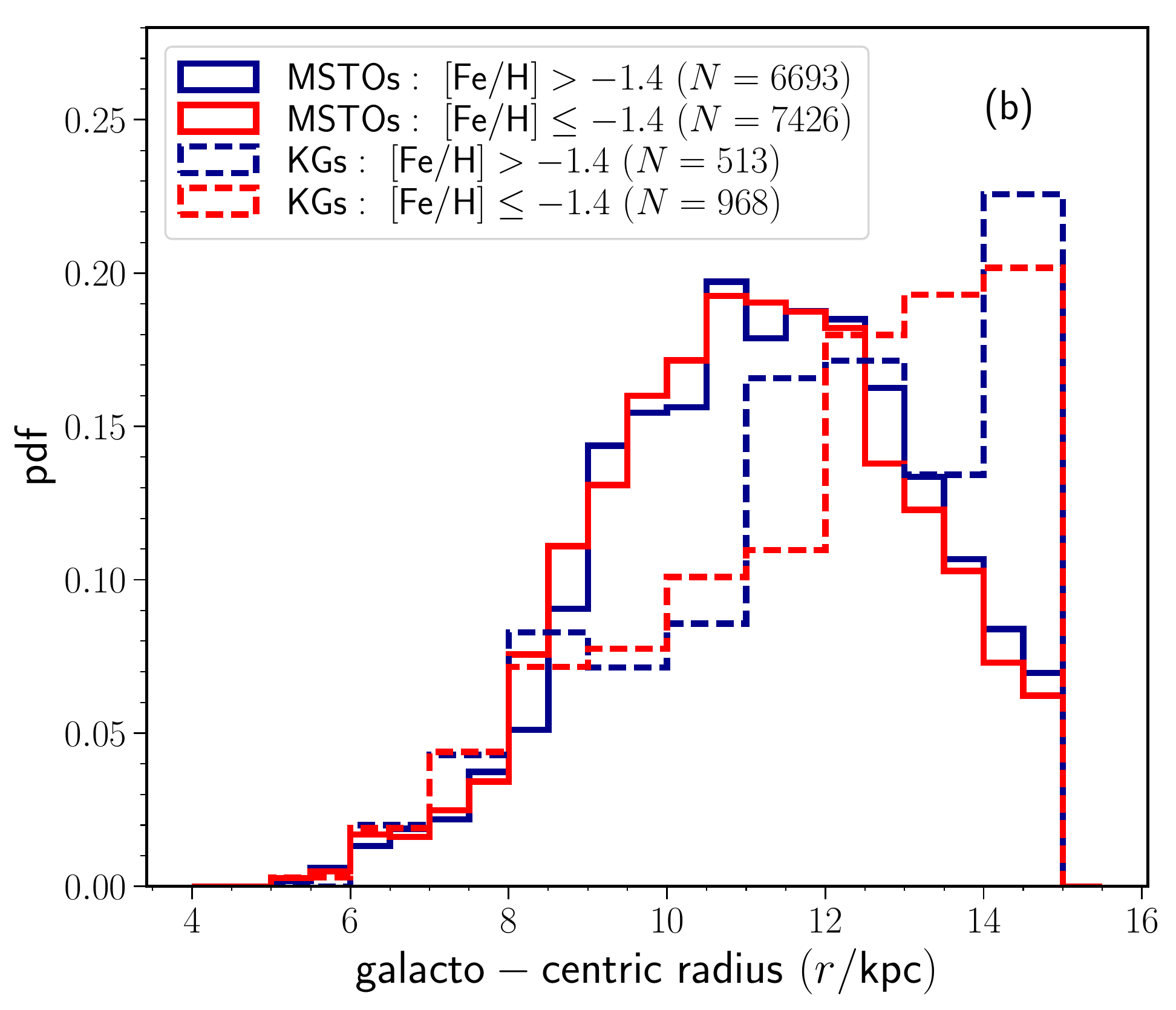}   
   \includegraphics[width=1.35\columnwidth]{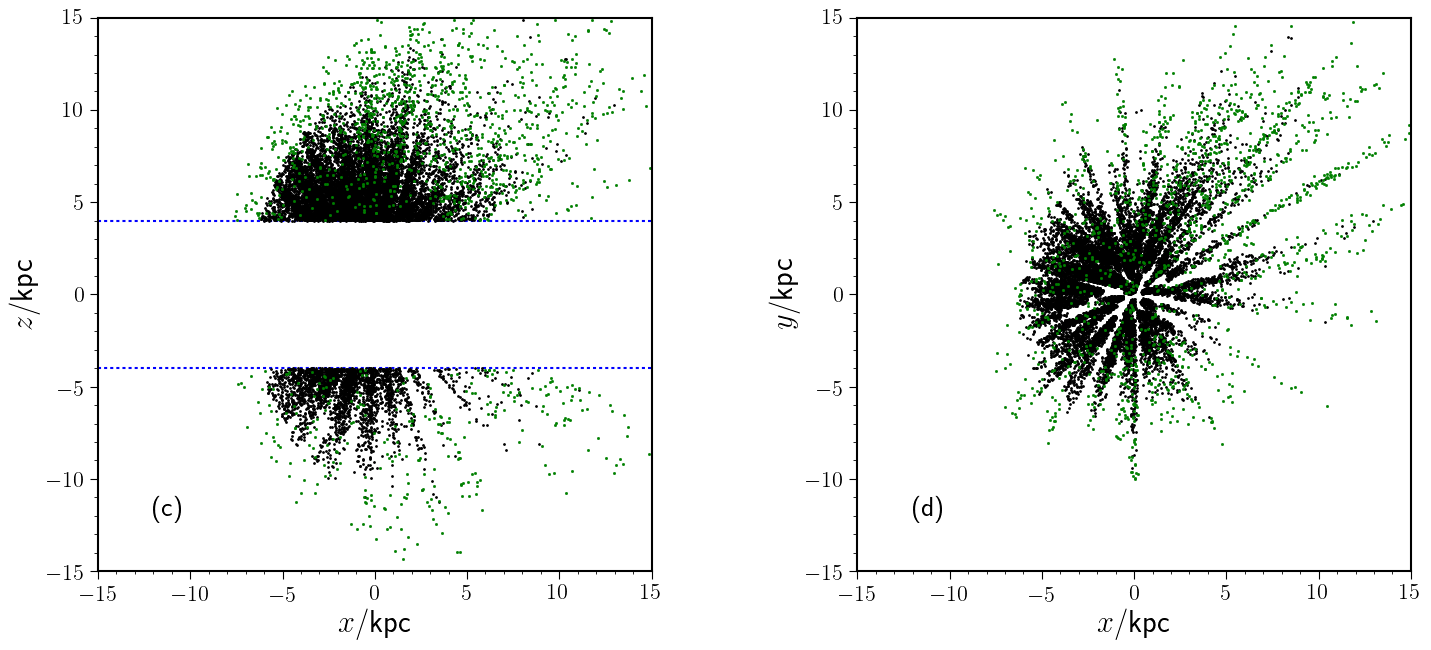}   
   \caption{Data properties of the SEGUE main-sequence turn-off (MSTO) and K-giant (KG) stars. 
            Panel (a) shows sky coverage of the stars used in this work in the galactic coordinates ($l,b$),
            panel (b) shows the number distribution of the stars in the stellar sub-samples in different [Fe/H] ranges and
            panels (c) and (d) show projection in cartesian coordinates.
            The blue dashed lines in panel (c) denotes |z|=4 kpc restriction we adopt to cull disk stars.}
    \label{fig:data}
\end{figure*}

We construct a MSTO dataset by querying SDSS DR13 \citep{2016arXiv160802013S} with the following colour and magnitude limits:
\begin{equation}\label{eqn:mstos}
\begin{cases}
0.15<g_0-r_0<0.4\\
3.5<\log g<4.7\\
14<r_0<20\\
\end{cases}
\end{equation}
where, $g_0$ and $r_0$ are the extinction corrected magnitudes and $g$ denotes the surface gravity.
The above colour restriction is similar to one used in \cite{2010AJ....140.1850B} and \cite{2011ApJ...728..106S}.
For each star in our initial sample, we calculate an absolute magnitude ($M_r$) using the following formula \citep{2008ApJ...684..287I}:
\begin{equation}\label{eqn:Mr}
\begin{split}
 \delta M_r  = &4.5 - 1.11 \mathrm{[Fe/H]} - 0.18 \mathrm{[Fe/H]}^2,\\ 
 M_r = & -5.06 + 14.32 (g_0-i_0) - 12.97 (g_0-i_0)^2 + \\
       & 6.127 (g_0-i_0)^3 - 1.267 (g_0-i_0)^4 + \\
       & 0.0967 (g_0-i_0)^5 + \delta M_r.
 \end{split}
\end{equation}
We expect some spread $\sigma_{M_r}$ around $M_r$, which we predict to increase as we go redward in $g_0-i_0$ colour and also as we go lower in [Fe/H]. 
To quantify this relation, we generate a synthetic catalogue of stars using \galaxia \footnote{\galaxia\ is a population synthesis code to creating a synthetic survey of the Milky Way based on its embedded model.
\galaxia\ uses isochrones from the Padova database to compute photometric magnitudes for the model stars \citep[][but see \cite{2004A&A...422..205G} for SDSS-specific details]{2008A&A...482..883M,1994A&AS..106..275B}.
The web-link to the \galaxia\ software is \url{http://galaxia.sourceforge.net/} and for the full documentation visit the web-link \url{http://galaxia.sourceforge.net/Galaxia3pub.html}.}\citep{2011ApJ...730....3S} 
from which we select MSTOs using the same selection function as above.
The dispersion in $\sigma_{M_r}$ as a function of the $g_0-i_0$ and [Fe/H] predicted by \galaxia\ is shown in the bottom panel of Fig.~\ref{fig:Mr}.
We interpolate $\sigma_{M_r}(g_0-i_0, {\rm [Fe/H]})$ relation shown above and then used it to compute the corresponding value
of $\sigma_{M_r}$ for SEGUE MSTOs given a $g_0-i_0$ colour and [Fe/H].

Likewise, using \galaxia\ we can also derive a $M_r(g_0-i_0, {\rm [Fe/H]})$ relation,
which is shown with the dashed lines in the top panel of Fig.~\ref{fig:Mr}.
But since the relation given in Equation~\ref{eqn:Mr} is known to be robust for SDSS stars \citep{2008ApJ...684..287I} we decide not to use the $M_r$ predicted by \galaxia\ in our final results. 
However, refer to Section~\ref{sec:test} for a comparison between distances estimated assuming two definitions of $M_r$ given above,
and also refer to Section~\ref{sec:sys} to understand the impact of this choice.

In the top panel of the Fig.~\ref{fig:Mr} solid lines demonstrate the colour and metallicity dependence of $M_r$ given by Equation~\ref{eqn:Mr}. 
With the estimated $M_r$ and associated dispersion $\sigma_{M_{r}}$, we calculate the distances to the MSTOs using the photometric parallax relation given by
\begin{equation}\label{eqn:dist}
s/\rm{kpc} = 10^{\mu/5-2},
\end{equation}
where $\mu=r_0-M_r$ is the distance modulus. 
For an uncertainty of $\sigma_{r_0}$ in the apparent magnitude ($r_0$) and $\delta M_r$ in the absolute magnitude ($M_r$) we calculate the uncertainty in the distance modulus $\sigma_\mu$ using,
\begin{equation}\label{eqn:muerr}
\sigma_\mu = \sqrt{\sigma_{M_{r}}^2 + \sigma_{r_0}^2}.
\end{equation}
This leads to a typical uncertainty of roughly $25\%$ in the estimated distances.
Note, when we use Equation~\ref{eqn:dist} to convert distance modulus $\mu$ to distance $s$, a Gaussian error function assumed for $\mu$ does not translate to a Gaussian error function for the distance.
Hence, in Section~\ref{sec:method} while measuring the kinematics we prefer to work in distance modulus space instead of distance.

Importantly, to ensure high fidelity we further impose quality cuts to only keep stars with 
$elodiervfinalerr>0$, $seguePrimary=1$ and $zwarning = 0 \rm{\ or\ } 16$ (taken from the SEGUE catalogue).
The SEGUE Stellar Parameter Pipeline \citep[SSPP,][]{2008AJ....136.2022L,2011AJ....141...89S} requires a S/N of at least 10 to estimate spectroscopic parameters, and the estimates tend to improve for higher S/N.
We hope that the quality (or lack thereof) of spectra reflects on the uncertainties quoted in the spectroscopic parameters, which we consider in our kinematic measurements (Section~\ref{sec:method}).
However, to avoid edge cases we only use stars with S/N>15.
Also, to select the potential halo stars we only use stars with the height above the galactic plane $|z|$ larger than $4\,\kpc$ as we wish to cull the disk contaminants.
Our ability to measure tangential velocity dispersions from $\vlos$ diminishes at larger radii \citep{2017arXiv170406286H}.
Moreover, we observe that uncertainties in distances increases at large radii. 
Therefore, in this paper we confine our study to the galacto-centric radius of $r/\kpc<15$.

Note, as recommended in the SEGUE documentation, we use \emph{elodiervfinal} for radial velocities,
\emph{elodiervfinalerr} for uncertainties in radial velocities and \emph{fehadop} for the [Fe/H] metallicity. 
In our final sample, $90\%$ of stars have $<20\%$ uncertainty in radial velocity.

\subsection{K-giant stars (KGs)}
We directly obtain KGs from the publicly available catalogue of \cite{2014ApJ...784..170X}, which is also constructed from the SEGUE project. 
These stars are selected based on $g_0-r_0$ and $u_0-g_0$ colours, and gravity. 
The distance to the stars are calculated using the photometric parallax relation for which they obtain an absolute magnitude by matching to a set of observables, 
i.e., colour, apparent magnitude and metallicity of a star to the metallicity dependent colour-absolute magnitude relations obtained from clusters. 
Typical uncertainty in the measured distance is claimed to be $16\%$.
Finally, to make this selection comparable to our MSTO sample, we impose $|z|/\kpc>4$ cut.
Compared to the MSTOs, KGs have better distance measurements, hence we relax the radial criteria and analyse KGs within $r/\kpc<17$.
This helps to increase the KGs sample size and allows us to determine comparatively more robust kinematics when using different metallicity ranges.

Finally, in Fig.~\ref{fig:data}(a) we show sky coverage of our selected MSTO and KG samples in galactic longitude ($l$) and latitude ($b$). 
Similarly, in Fig.~\ref{fig:data}(b) we present the number distribution of our stellar samples as a function of galacto-centric radius $r$ for different metallicity ranges as labelled in the figure.
The metallicity division for both the MSTO and KG samples is defined at the median [Fe/H] of the full sample.
We test that a shift of 0.2 dex to the division in either direction does not change our conclusions.
In the figure the comparatively metal-rich ([Fe/H]$>-1.4$) and the metal-poor ([Fe/H]$\leq-1.4$) MSTOs/KGs are shown in blue and red colours respectively,
where the solid line represents the MSTOs and the dashed line represents the KGs.
As labelled in the figure, the number of MSTOs with [Fe/H]$>-1.4$, MSTOs with [Fe/H]$\leq-1.4$, KGs with [Fe/H]$>-1.4$ and KGs with [Fe/H]$\leq-1.4$ are 6693, 7426, 513 and 968 respectively.
We find that the metal-rich and metal-poor MSTOs or KGs have roughly similar radial distributions.
Additionally, in panels (c) and (d) we show spatial distributions of our final MSTOs (black dots) and KGs (green dots) samples. 
The dashed blue lines in panel (c) shows |z|=4 kpc demarcation; the spatial limit applied to cull disk stars.
The radial spikes seen in panel (c) and (d) are the natural feature of the pencil-beam observation made by the SEGUE survey.

\section{Method}\label{sec:method}
\begin{figure*}
   \centering
   \includegraphics[width=1.\columnwidth]{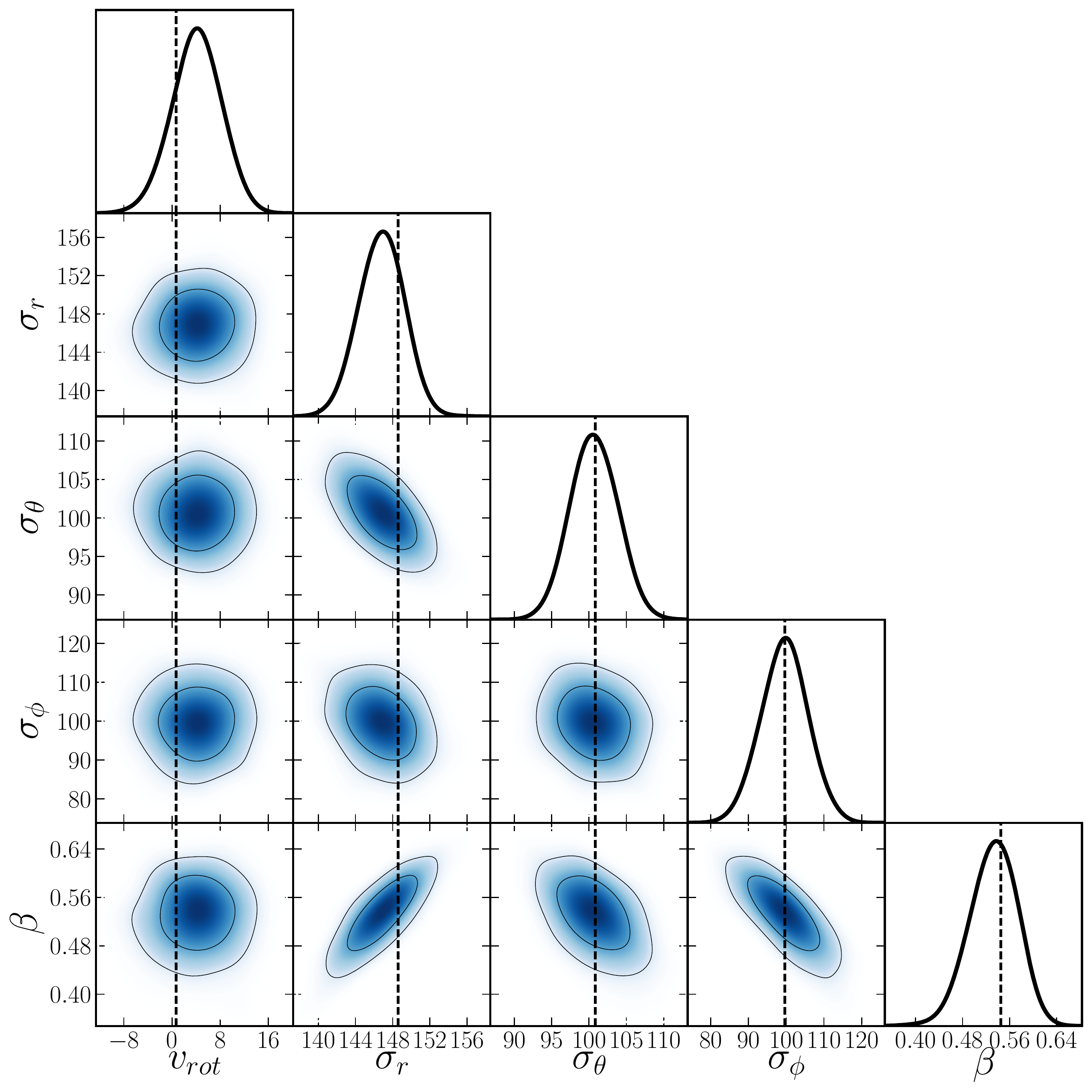}
   \includegraphics[width=1.\columnwidth]{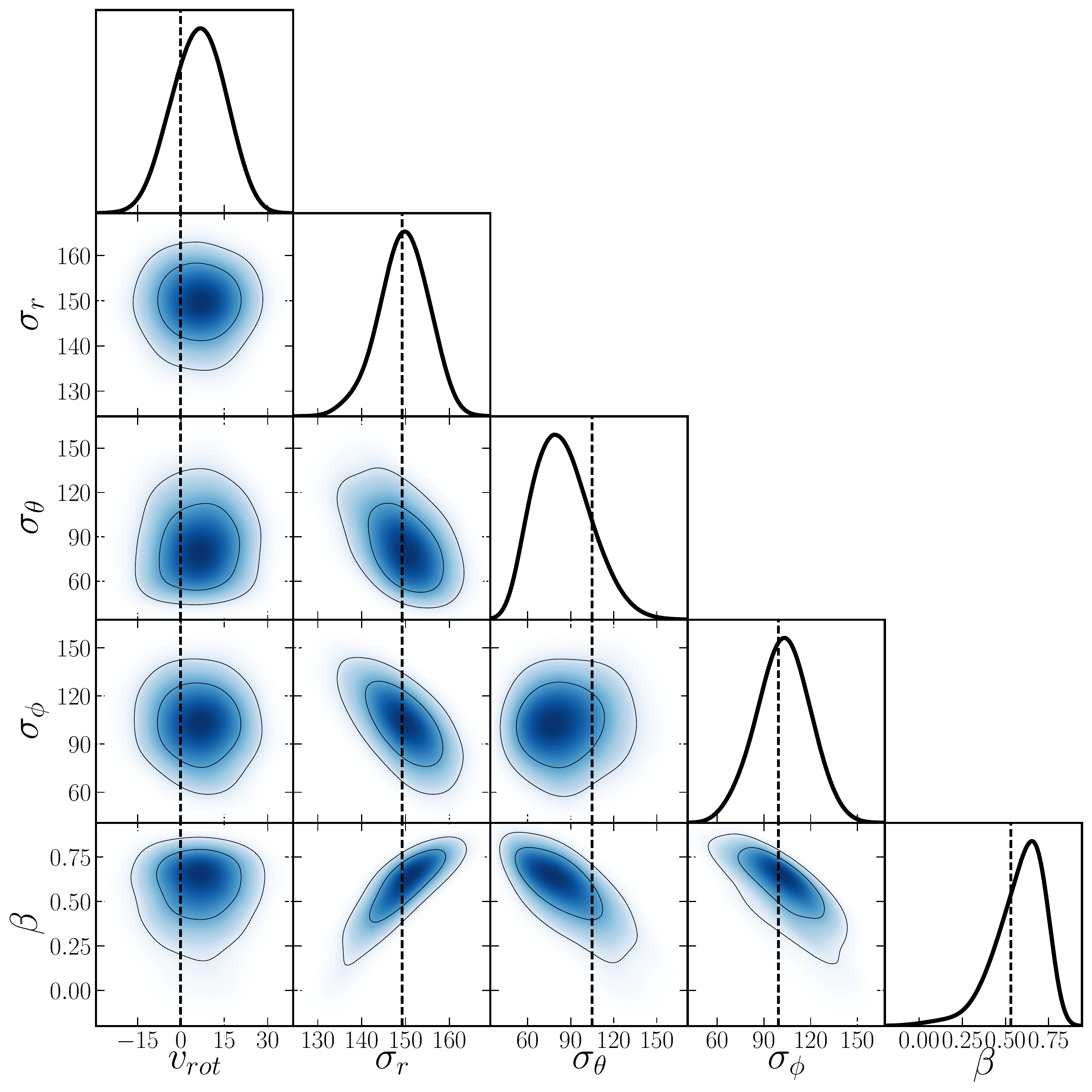}
   \caption{Results of the test runs on the metal-poor samples of MSTOs (left panel) and KGs (right panel) with radially biased synthetic velocity distributions.
            The contours show joint posterior probability distributions of the velocity moments for the synthetic data and the histograms show one dimensional marginalised posterior distribution.
            The synthetic data has same number of stars and sky footprint as the SEGUE data, 
            in fact it uses exactly same values of $l, b, \mu$ and $\sigma_{v_{\rm los}}$ as the SEGUE data. 
            Dashed vertical lines show the true values of the velocity moments of the synthetic data.
            The contours depict $1\sigma$ and $2\sigma$ credibility intervals.}
    \label{fig:testcorner}
\end{figure*}
From our heliocentric vantage point in the Milky Way we can observe the angular positions on the sky (i.e., Galactic longitude $l$ and latitude $b$), distance ($s$), 
line-of-sight velocity ($\vlos$), and proper motion (tangential motion on the sky $\mu_l$ and $\mu_b$) for each star, albeit with some uncertainty.
Given the distance of the Sun ($R_0$) from the Galactic centre, its relative motion with respect to the Local Standard of Rest (LSR) of the Galaxy ($U_\odot$, $V_\odot$, $W_\odot$)
and the motion of the LSR with respect to the Galactic centre ($v_{\text LSR}$), we can convert
the heliocentric coordinates to the Galactocentric reference frame according to Appendix~\ref{sec:coordtransform}. 
The velocity distribution of halo stars in the Galactocentric reference frame can then be described by a multivariate Gaussian model of the form:
\begin{equation}\label{eqn:velelp}
\begin{split}
p(v_r, v_\theta, v_\phi|\Theta, l, b, s(\mu)) \propto & {\cal N}(v_r|0, \rsigma) {\cal N}(v_\theta|0, \thsigma)\\
& {\cal N}(v_\phi|v_{\text rot}, \phsigma),
\end{split}
\end{equation}
where ${\cal N}$(.|mean, variance) refers to a Normal distribution, and
$\Theta = \{ \rsigma, \thsigma, \phsigma, v_{\text rot} \}$ represents a set of parameters that 
include the mean azimuthal velocity ($v_{\text rot} = \langle v_\phi \rangle$) and the velocity dispersion profiles in spherical coordinates along the radial ($r$), angular ($\theta$) and azimuthal ($\phi$) directions.
Note, here a positive value for $v_{\text{rot}}$ means retrograde motion whereas a negative value means prograde motion i.e. rotating in the same direction the Galactic disk rotates. 
Following the recent works of \cite{2009ApJ...698.1110S}, \cite{2010ApJ...716....1B}, \cite{2015ApJ...813...89K} and \cite{2016MNRAS.460.1725D}
we assume that the velocity ellipsoids of the halo stars are aligned along the coordinate frame directions of the spherical polar coordinate centred at the Galactic centre.
As a consequence, the covariance matrix of the velocity ellipsoid is assumed to be diagonal i.e. we ignore correlations among radial, angular and azimuthal velocities. 
In general, tangential velocities ($v_l$ and $v_b$ or proportionally proper-motion) 
of the stellar halo stars are either completely unknown or not known accurately.
But, the distance modulus and line-of-sight velocities are known with uncertainties of $\sigma_\mu$ and $\sigma_{\vlos}$ respectively. 
The true, uncertainty-free version of the distance modulus $\mu'$ and line-of-sight velocity $\vlos'$ are not accessible and we treat these as hidden/latent variables. 
We marginalise over the unknowns-- tangential velocities, true distance modulus and true line-of-sight velocities-- to obtain 
\begin{equation}\label{eqn:likelihood}
  \begin{split}
  p(\vlos|\Theta, l, b, \mu, \sigma_\mu, \sigma_{\text los}) = \iiiint p(v_l, v_b, \vlos|\Theta,l,b,\mu',\vlos')\\
  p(\vlos'|\vlos, \sigma_{\vlos})~p(\mu'|\mu, \sigma_\mu)~p(\mu')~{\text d} v_l~{\text d} v_b~{\text d}\vlos'~{\text d}\mu'.
  \end{split}
\end{equation}
To determine the distance modulus distribution $p(\mu)$ we assume that 
$p(\mu)~{\text d}\mu = \rho(l,b,s) 4\pi s^2 {\text d} s$ and utilise Equation~\ref{eqn:dist} to derive
\begin{equation}\label{eqn:muprior}
p(\mu) \propto s(\mu)^3 \rho(l,b,s(\mu)).
\end{equation}
Here, $\rho(l,b,s(\mu))$ is a galacto-centric radial distribution of the halo stars.
Fortunately, from recent observational evidence such as
\cite{2008ApJ...680..295B,2009MNRAS.398.1757W,2011ApJ...731....4S,2011MNRAS.416.2903D,2012ApJ...756...23A,2015ApJ...809..144X}
it is known that the logarithmic density distribution of the smooth component of the inner stellar halo follow a power-law given by 
$\rho \propto r(l,b,s(\mu),R_0)^{-\alpha}$ with power-law slope $\alpha\simeq2.5$.
However, in Section~\ref{sec:results} we investigate the effect of our choice of power-law slope on our final results.

Finally, given the data D$=\{l,b,\mu \pm \sigma_\mu, \vlos \pm \sigma_{\vlos}\}$ of N stars we estimate $\Theta$ by
\begin{equation}\label{eqn:posterior}
p(\Theta|D) \propto p(\Theta) \prod_i^N p(\vlos|\Theta, l, b, s, \sigma_s, \sigma_{\text los}), 
\end{equation}
where $p(\Theta)$ represent the priors on model parameters $\Theta$.
Eqn.~\ref{eqn:likelihood} does not have an analytical form. 
Hence, marginalisation needs to be done numerically using some deterministic integration techniques, which is inefficient. 
Alternatively, we can treat the missing data or unknowns, $v_l, v_b, s'$ and $\vlos'$, as latent variables by setting them up as a hierarchical Bayesian model and estimating them simultaneously alongside $\Theta$
using the Metropolis-within-Gibbs algorithm \citep[][\citealt{2017arXiv170601629S}\footnote{the code is available at \url{https://github.com/sanjibs/bmcmc}}]{2014ApJ...793...51S}.
To ensure the stability of the distributions of parameters around certain values, we run the algorithm for a sufficient autocorrelation time.
We consider the values corresponding to the median of the posterior distributions of the 
parameters $\Theta$ as the best estimates, and the $16 ^{\text{th}}$ and $84^{\text{th}}$ percentiles of the distributions as its associated uncertainties. 

We choose flat priors for  
$(\rsigma, \thsigma, \phsigma)/\kms \in [10,250]$; $\vrot/\kms \in [-60,60]$ and $(v_l, v_b)/\kms \in [-600, 600]$.
We assume $R_0 = 8.2$ kpc \citep{2016ARA&A..54..529B}, $U_\odot=11.1\kms$, $V_\odot=12.24\kms$, $W_\odot=7.25\kms$ \citep{2010MNRAS.403.1829S} 
and $v_{\text{LSR}} = 236.2\kms$ \citep{2004ApJ...616..872R}.
However, later in the text we investigate and discuss the implications of our choice of priors on our final results.

Finally, we would like to highlight that our method also has an immediate application for the upcoming data releases of \emph{Gaia} 
even if it could not provide a reliable radial velocity for halo stars. 
In such a case we can marginalise over $\vlos$ and utilise the observed parallax distances, proper-motion and associated uncertainties instead.

\section{Tests with simulated data sets}\label{sec:test}
There are two important aspects of our analysis that need to be tested. 
First, the efficacy of our marginalisation technique (Equation~\ref{eqn:likelihood}) 
and second, the effect of observational biases and variance in our final results.

For the first aspect we perform a simple test, where we construct a set of synthetic data that mimic spatial distributions of our MSTO/KG sample.
Here we adopt the original position vector $(l,b,\mu \pm \sigma_\mu)$ of the SEGUE MSTOs/KGs sample.
This way we preserve the effect of the survey footprint and sample size in the test analysis identical to our final analysis.
We then sample Equation~\ref{eqn:velelp} to obtain synthetic velocities, which we project to the observational space and replace the observed SEGUE line-of-sight velocity, $\vlos$, with the synthetic $\vlos$. 
To realise the error distribution of $\vlos$ we pair the sampled, $\vlos$, with the actual, $\sigma_{\rm los}$, of the SEGUE stars.
Later we analyse the observed data in an identical manner.
First, in the left panel of Fig.~\ref{fig:testcorner}, we show a case of the SEGUE MSTOs sample with radial velocity anisotropy. 
The contours in the figure show the $1\sigma$ and $2\sigma$ regions of the joint posterior probability distributions of the model parameters, as well as the derived velocity anisotropy $\beta$, sampled from the MCMC runs. 
The input value of each parameter in this example were $\rsigma=150~\kms, \thsigma=\phsigma=100~\kms$ and $\vrot=0~\kms$, which could be slightly altered while sub-sampling due to the stochastic nature of the randomness trials.  
The true values are shown with the vertical dashed lines in the figure. 
Similarly, in the right panel of Fig.~\ref{fig:testcorner} we also show a case of the SEGUE KGs sample, with the same values for the model parameters as for the MSTO case.
We observe that the intrinsic velocity moments in both the above mentioned cases are recovered within the $1\sigma$ credibility interval. 
We also repeat the exercise with possible alternative scenarios for example assuming isotropic/tangential velocity anisotropies for MSTOs/KGs and also, for positive and negative values for $\vrot$.
Furthermore, we also repeat the above set of tests with metal-rich subsets of MSTOs/KGs.
In all of these additional cases we achieve similar levels of accuracy as demonstrated in Fig.~\ref{fig:testcorner}.

Now for the second part we perform a more realistic test, for which we invoke \galaxia. 
Here we aim to understand effects of observational errors that are mainly due to sample selection, survey footprint, distance estimations and random errors in the observables on the final kinematics.
For this we first generate synthetic MSTO stars using \galaxia\ with a more relaxed selection criteria 
(to allow margin for uncertainties in each parameter) on $r_0$ and $g_0-r_0$ than the one provided in Equation~\ref{eqn:mstos}.
Then we jitter the colour ($g_0-r_0$), and also the surface gravity ($\log g$) and the metallicity ([Fe/H])
for each star assuming Gaussian error distributions with dispersions equal to 0.03 mag, 0.28 dex and 0.20 dex respectively 
\footnote{We adopt uncertainties in stellar and photometric parameters of the SEGUE stars from \url{http://www.sdss3.org/dr9/spectro/sspp_internal.php}}.
Now, we impose exact selection criteria provided in Equation~\ref{eqn:mstos} to the synthetic catalogue.
Furthermore, to mimic SEGUE footprint shown in Figure~\ref{fig:data}(a)
we employ HEALPix \citep{1999astro.ph..5275G,2005ApJ...622..759G}(with npixel=768) 
and only select stars in the pixels that overlap with the SEGUE footprint.
\begin{figure}
   \centering
   \includegraphics[width=1.0\columnwidth]{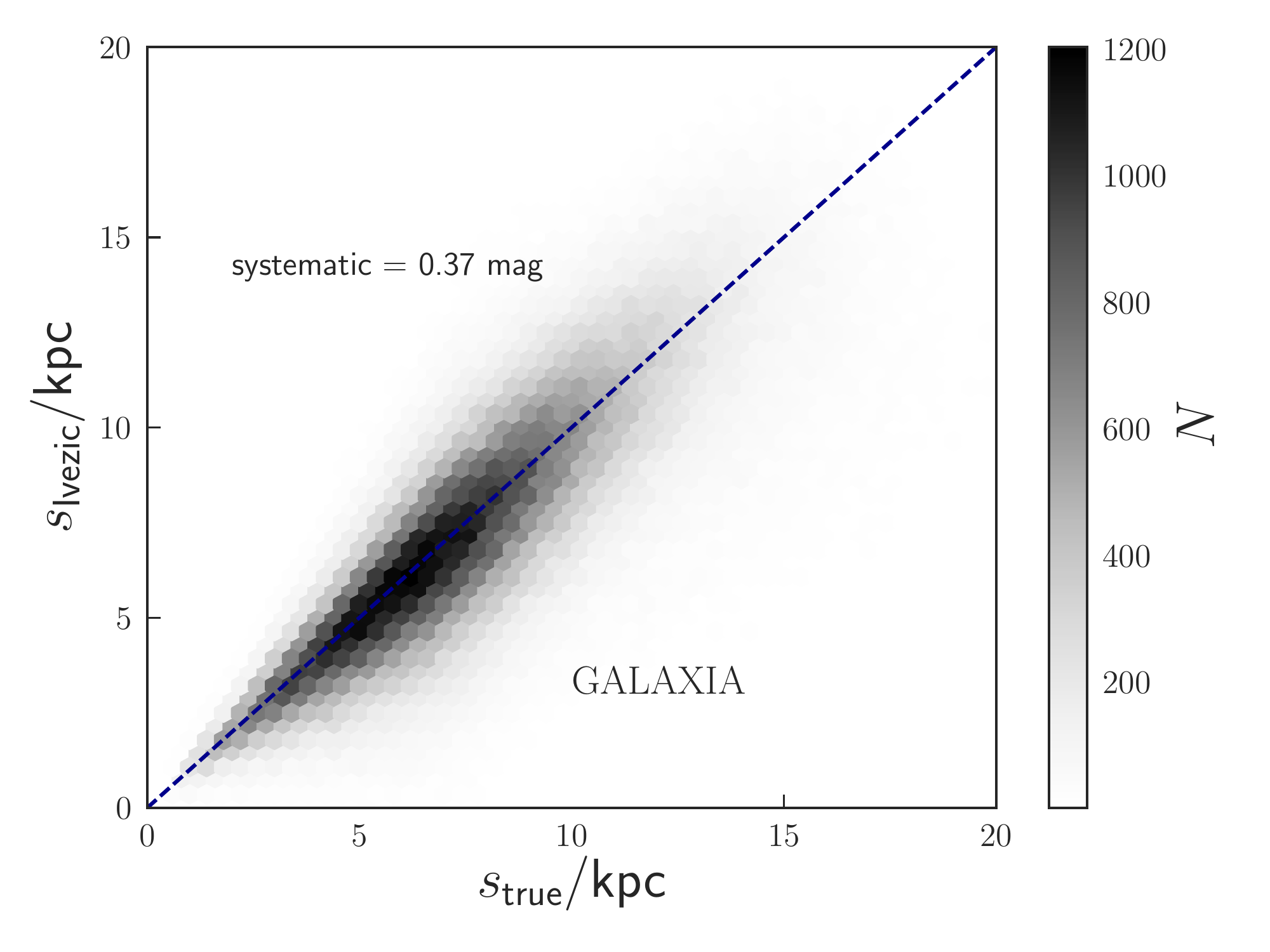}   
   \caption{A comparison between distances estimated using \protect\cite{2008ApJ...684..287I} relation and intrinsic distances for synthetic MSTO sample obtained from \galaxia.}
   \label{fig:dist_comparison}
\end{figure}
\begin{figure}
   \centering
   \includegraphics[width=1.\columnwidth]{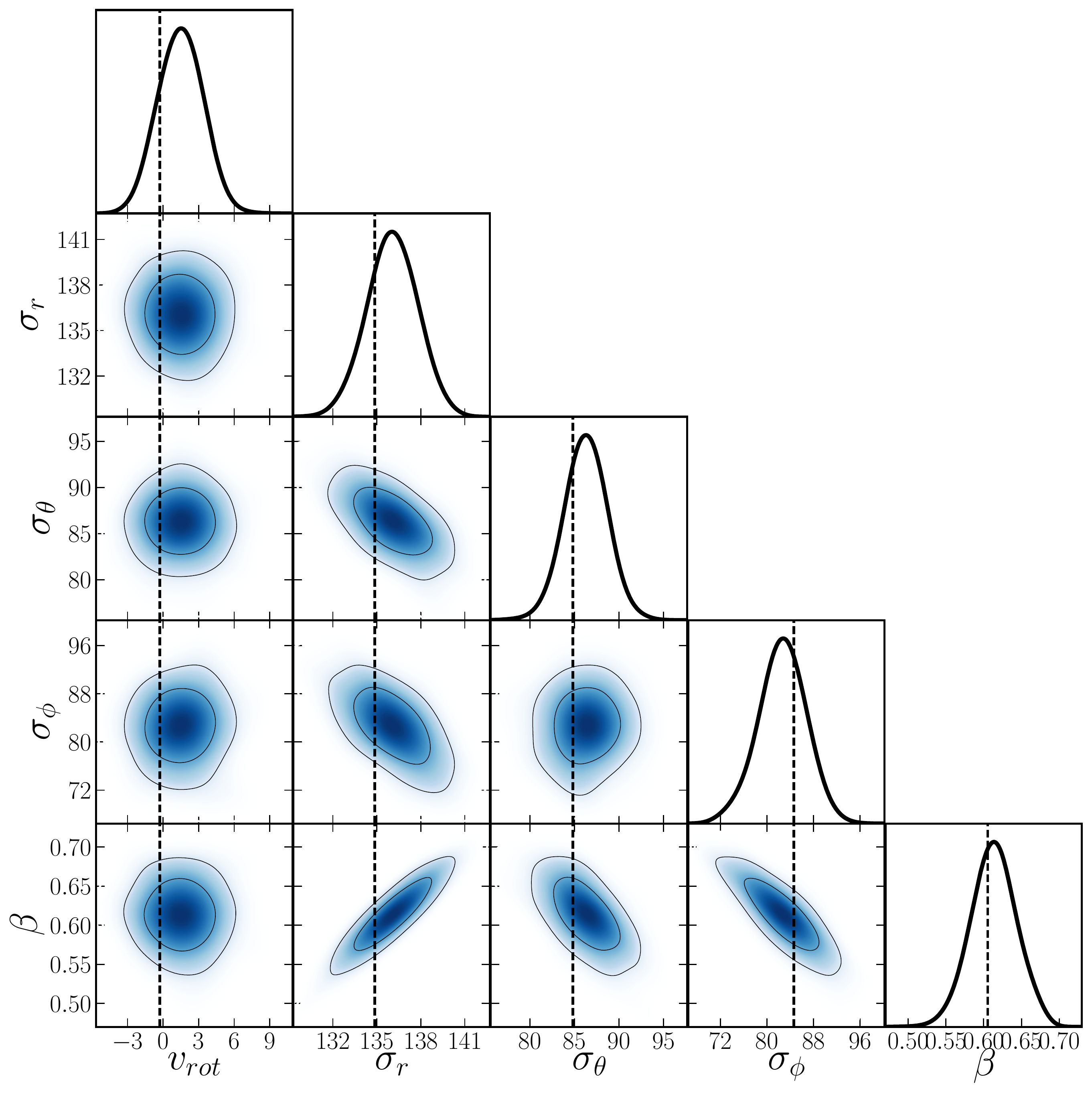}
   \caption{Results of the test runs on the synthetic MSTOs sampled from \galaxia\ and convolved with SEGUE like error functions.
            The contours show joint posterior probability distributions of the velocity moments for the synthetic data and the histograms show one dimensional marginalised posterior distribution.
            The synthetic data has same number of stars and roughly same sky footprint as the SEGUE data.
            Moreover, they are convolved with SEGUE like uncertainties in photometry, stellar parameters and radial velocity. 
            Dashed vertical lines show the true values of the velocity moments of the synthetic data.
            The contours depict $1\sigma$ and $2\sigma$ credibility intervals.}
    \label{fig:testcornergalaxia}
\end{figure}
Next, we use the \cite{2008ApJ...684..287I} relation given in Equation~\ref{eqn:Mr} to derive $M_r$ for the synthetic MSTO
as we did for the SEGUE MSTOs in Section~\ref{sec:data}.
We find that $M_r$ derived from \galaxia\ is systematically higher by an average +0.37 mag compared to the \cite{2008ApJ...684..287I} relation.
This could be due to zero-point offsets in SDSS magnitudes provided in the isochrones. 
Similarly, we calculate dispersion from the interpolated function $\sigma_{M_r}(g_0-i_0, {\rm [Fe/H]})$ derived in Section~\ref{sec:data}.
Then we use Equation ~\ref{eqn:dist} to calculate distances to the synthetic MSTOs.  
In Fig.~\ref{fig:dist_comparison} we compare the estimated distances for the synthetic MSTOs against their intrinsic distances provided by \galaxia.
The blue dashed line in the figure shows 1:1 correspondence between the two distances whereas the colour code in the figure represents the number of stars in each pixel.
We can see that the two distances are in reasonable agreement, considering the systematic as well as given the realistic (observational-like) errors.
To avoid confusion, we would like to re-emphasise here that the \cite{2008ApJ...684..287I} relation can be assumed to be robust for the SDSS/SEGUE stars, hence needs no such correction. 
Rather, to mock SEGUE stars, $M_r$ of the \galaxia\ MSTOs need to be corrected by $+0.37$ mag.
Utilising the estimated distances and angular positions we derive the galacto-centric position vector of the tracers. 
Then we impose spatial cuts of $r<15\,\kpc$ and $|z|<4\,\kpc$ to select the halo stars, as we do in the case of the SEGUE MSTOs.
In the end we also add a Gaussian random error to the radial velocities of the synthetic MSTOs,
where the error distribution is kept the same as the error distribution of the radial velocities of the observed data (SEGUE MSTOs).
Finally, we feed processed synthetic MSTOs data D=$\{l,b,\mu \pm \sigma_{\mu}, \vlos \pm \sigma_{\vlos} \}$ 
to our machinery (Section~\ref{sec:method}) to derive the model parameters.
Joint probabilities of the model parameters obtained from this exercise are shown in Fig.~\ref{fig:testcornergalaxia}.
Dashed vertical lines at $\rsigma=134.9\,\kms, \thsigma=84.8\,\kms, \phsigma=84.6\,\kms, \vrot=-0.3\,\kms)$ shows the intrinsic values of the velocity moments of the synthetic MSTOs.
We can see that the true values of the moments fall within $1\sigma$ regions of the estimated kinematics.
The above tests demonstrate that in the radial range of our data $r\in[0,15]\,\kpc$, 
$\vlos$ information is sufficient to recover the full kinematics $(\rsigma, \thsigma, \phsigma, \vrot)$ 
of the halo, even when the observational biases/variances are taken into account.

\section{Results}\label{sec:results}
\subsection{Kinematics of the Turn-off and K-giant stars}\label{sec:resultsmstokg}
\begin{table*}
\begin{minipage}{150mm}
\caption{Kinematics of the inner stellar halo of the Milky Way.}
\label{table:vmoments}
\begin{tabular}{@{}lcccccccc}
\hline
Stellar & Cases & Number of& $\vrot$ & $\sigma_r$ & $\sigma_\theta$ & $\sigma_\phi$ & $\beta$\\ 
population & & stars &(km\,s$^{-1}$)& (km\,s$^{-1}$)& (km\,s$^{-1}$)& (km\,s$^{-1}$)&  \\ 
\hline
\hline
\multicolumn{7}{|c|}{Main-sequence turn-off (MSTO) and K-giant (KG) stars with $|z|/\kpc>4$, and in different metallicity bins} \\ \hline
MSTOs & [Fe/H]$\leqslant-1.4, r/\kpc \leqslant15$ & 7426 & $26^{+4}_{-4}$ & $141^{+3}_{-2}$ & $109^{+4}_{-4}$ & $82^{+8}_{-8}$ & $0.53^{+0.05}_{-0.05}$ \\ 
MSTOs & [Fe/H]$>-1.4, r/\kpc \leqslant 15$ & 6693 & $-16^{+4}_{-4}$ & $129^{+3}_{-2}$ & $93^{+4}_{-4}$ & $72^{+9}_{-9}$ & $0.58^{+0.06}_{-0.05}$ \\ 
KGs & [Fe/H]$\leqslant-1.4, r/\kpc \leqslant 17$ & 968 & $-7^{+8}_{-8}$ & $142^{+5}_{-5}$ & $70^{+10}_{-10}$ & $110^{+10}_{-10}$ & $0.56^{+0.1}_{-0.1}$ \\ 
KGs & [Fe/H]$>-1.4, r/\kpc \leqslant 17$ & 513 & $-20^{+10}_{-10}$ & $140^{+7}_{-7}$ & $90^{+20}_{-20}$ & $100^{+20}_{-20}$ & $0.5^{+0.1}_{-0.1}$ \\ 
\hline
\multicolumn{7}{|c|}{Metal-poor ([Fe/H]$\leqslant-1.4$) MSTOs in $r$ and $|z|$ bins} \\ \hline
MSTOs & $|z|/\kpc>4,r/\kpc \leqslant 11$ & 3495 & $20^{+5}_{-5}$ & $139^{+6}_{-6}$ & $108^{+5}_{-5}$ & $90^{+10}_{-10}$ & $0.48^{+0.1}_{-0.09}$ \\ 
MSTOs & $|z|/\kpc>4,11<r/\kpc \leqslant 15$ & 3930 & $40^{+6}_{-6}$ & $142^{+4}_{-4}$ & $110^{+10}_{-10}$ & $60^{+8}_{-8}$ & $0.62^{+0.09}_{-0.08}$ \\ 
MSTOs & $r/\kpc \leqslant 15, 4<|z|/\kpc<5.3$ & 3677 & $21^{+5}_{-5}$ & $136^{+3}_{-3}$ & $103^{+4}_{-4}$ & $84^{+9}_{-9}$ & $0.52^{+0.06}_{-0.06}$ \\ 
MSTOs & $r/\kpc \leqslant 15, |z|/\kpc> 5.3$ & 3748 & $34^{+6}_{-6}$ & $142^{+4}_{-4}$ & $118^{+6}_{-6}$ & $80^{+10}_{-10}$ & $0.49^{+0.08}_{-0.08}$ \\ 
MSTOs & $|z|/\kpc>5.3, r/\kpc \leqslant 11$ & 1362 & $23^{+8}_{-8}$ & $140^{+10}_{-10}$ & $121^{+9}_{-9}$ & $100^{+20}_{-20}$ & $0.3^{+0.2}_{-0.2}$ \\ 
\hline
\multicolumn{7}{|c|}{Metal-rich ([Fe/H]$>-1.4$) MSTOs in $r$ and $|z|$ bins} \\ \hline
MSTOs & $|z|/\kpc>4,r/\kpc \leqslant 11$ & 2985 & $-23^{+5}_{-5}$ & $135^{+5}_{-5}$ & $84^{+6}_{-6}$ & $70^{+10}_{-10}$ & $0.67^{+0.07}_{-0.07}$ \\ 
MSTOs & $|z|/\kpc>4,11<r/\kpc \leqslant 15$ & 3705 & $1^{+7}_{-7}$ & $122^{+4}_{-3}$ & $114^{+6}_{-9}$ & $70^{+10}_{-10}$ & $0.4^{+0.1}_{-0.1}$ \\ 
MSTOs & $r/\kpc \leqslant 15, 4<|z|/\kpc<5.3$ & 3472 & $-30^{+5}_{-5}$ & $120^{+3}_{-3}$ & $85^{+5}_{-5}$ & $83^{+8}_{-8}$ & $0.51^{+0.07}_{-0.07}$ \\ 
MSTOs & $r/\kpc \leqslant 15, |z|/\kpc> 5.3$ & 3218 & $6^{+6}_{-6}$ & $133^{+4}_{-4}$ & $107^{+7}_{-7}$ & $70^{+10}_{-10}$ & $0.54^{+0.09}_{-0.09}$ \\ 
MSTOs & $|z|/\kpc>5.3, r/\kpc \leqslant 11$ & 1116 & $-1^{+9}_{-9}$ & $134^{+9}_{-9}$ & $104^{+10}_{-10}$ & $70^{+10}_{-10}$ & $0.5^{+0.1}_{-0.1}$ \\ 
\hline
\multicolumn{7}{|c|}{MSTOs $(|z|/\kpc>4,r/\kpc \leqslant15)$ with systematics} \\ \hline
MSTOs & $\vlos +5 \,\kms$, [Fe/H]$\leqslant-1.4$ & 7426 & $22^{+4}_{-4}$ & $141^{+3}_{-3}$ & $109^{+4}_{-4}$ & $82^{+8}_{-8}$ & $0.53^{+0.05}_{-0.05}$ \\ 
MSTOs & $\galaxia\ M_r\,\rm{calibration}$, [Fe/H]$\leqslant-1.4$ & 5426 & $32^{+5}_{-5}$ & $136^{+3}_{-3}$ & $121^{+4}_{-4}$ & $88^{+9}_{-9}$ & $0.39^{+0.07}_{-0.06}$ \\ 
MSTOs & $\galaxia\ M_r\,\rm{calibration}$, [Fe/H]$>-1.4$ & 4879 & $-2^{+5}_{-5}$ & $128^{+3}_{-3}$ & $109^{+4}_{-4}$ & $88^{+9}_{-9}$ & $0.39^{+0.08}_{-0.08}$ \\ 
\hline
\end{tabular} 
\medskip
$\rsigma, \thsigma, \phsigma$ are the velocity dispersions in spherical coordinates, $\beta$ is the velocity anisotropy parameter and $v_{\rm rot}$ is the mean azimuthal velocity; measured in the galacto-centric reference frame. A positive value for $v_{\rm rot}$ means retrograde motion whereas a negative value means prograde motion i.e. rotating in the same direction Galactic disk rotates.
\end{minipage}
\end{table*}
\begin{figure}
   \centering
   \includegraphics[width=1\columnwidth]{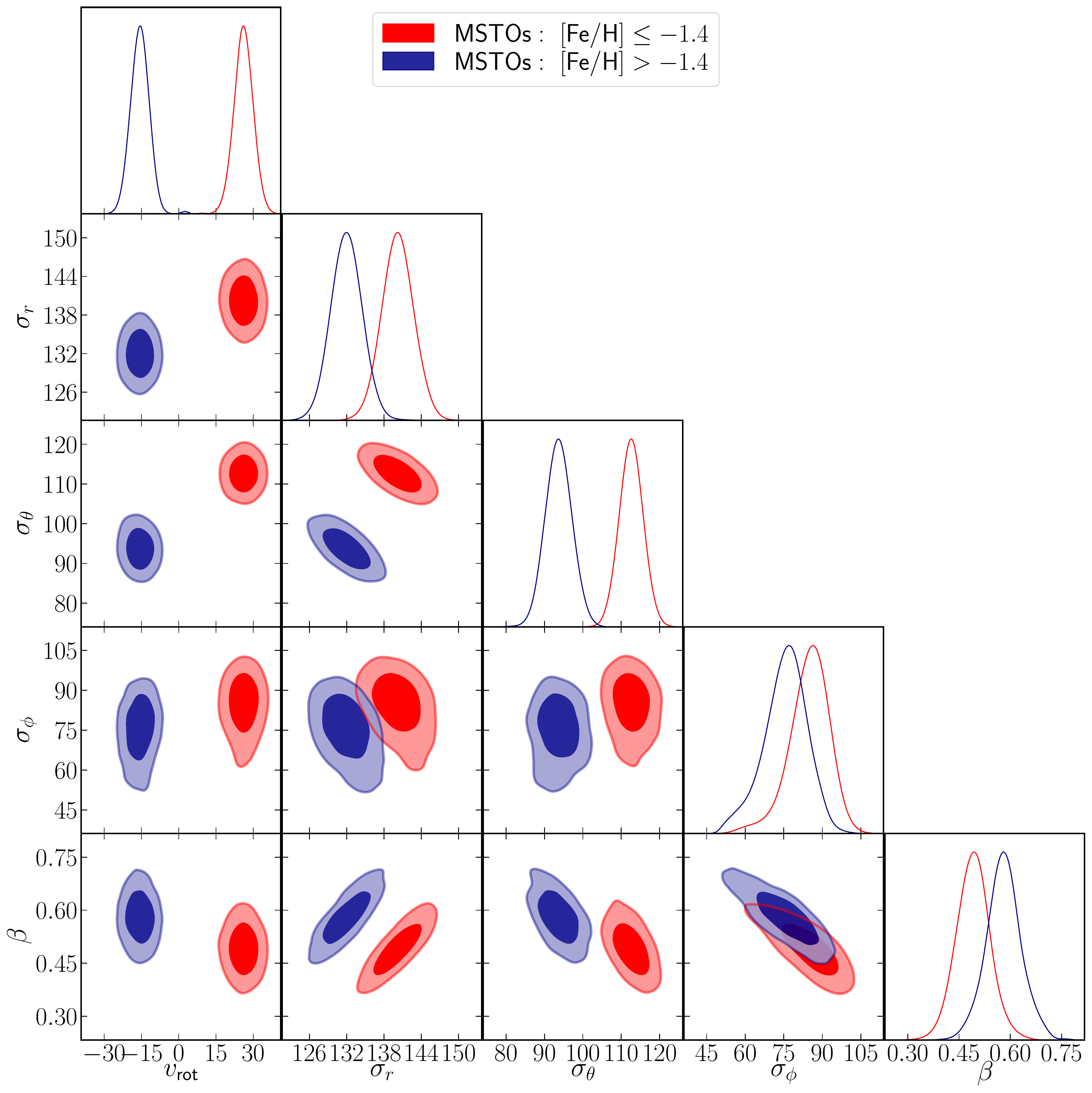}
   \caption{Kinematics of main-sequence turn-off stars (MSTOs). 
            The contours show joint posterior probability distributions of the velocity moments and anisotropy to $1\sigma$ and $2\sigma$ credibility intervals
            and the histograms show one dimensional marginalised posterior distribution.
            The blue and red colours show posteriori for the comparatively metal-rich (\feh$>-1.4$) and metal-poor (\feh$\leqslant-1.4$) sub-sample of the MSTOs respectively.}
    \label{fig:mstocorner}
\end{figure}
\begin{figure}
   \centering
   \includegraphics[width=1\columnwidth]{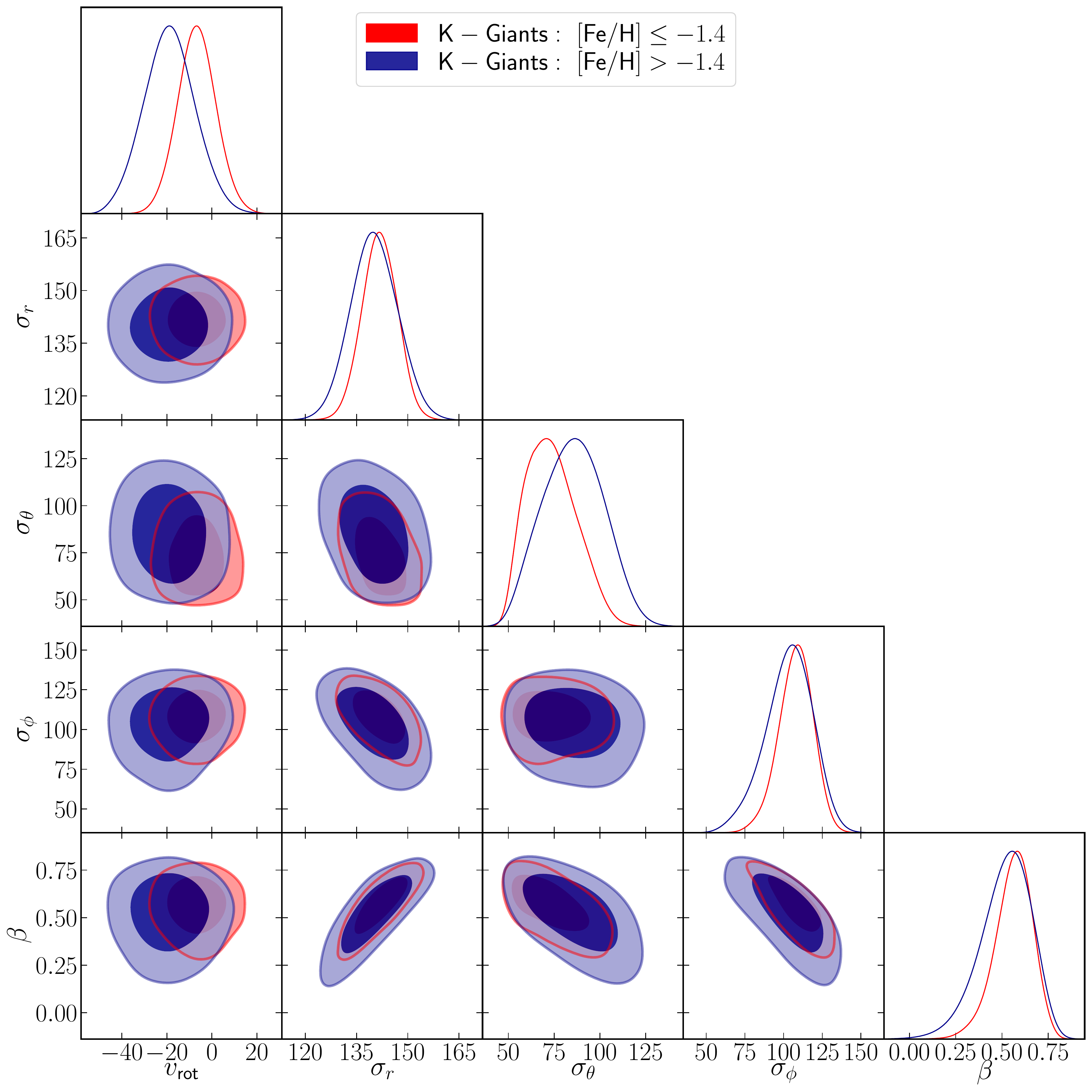}
   \caption{Kinematics of K-giant stars (KG). 
            The contours show joint posterior probability distributions of the velocity moments and anisotropy to $1\sigma$ and $2\sigma$ credibility intervals
            and the histograms show one dimensional marginalised posterior distribution.
            The blue and red colours show posteriori for the comparatively metal-rich (\feh$>-1.4$) and metal-poor (\feh$\leqslant-1.4$) sub-sample of the KGs respectively.}
    \label{fig:kgcorner}
\end{figure}
\begin{figure*}
   \centering
   \includegraphics[width=0.8\columnwidth]{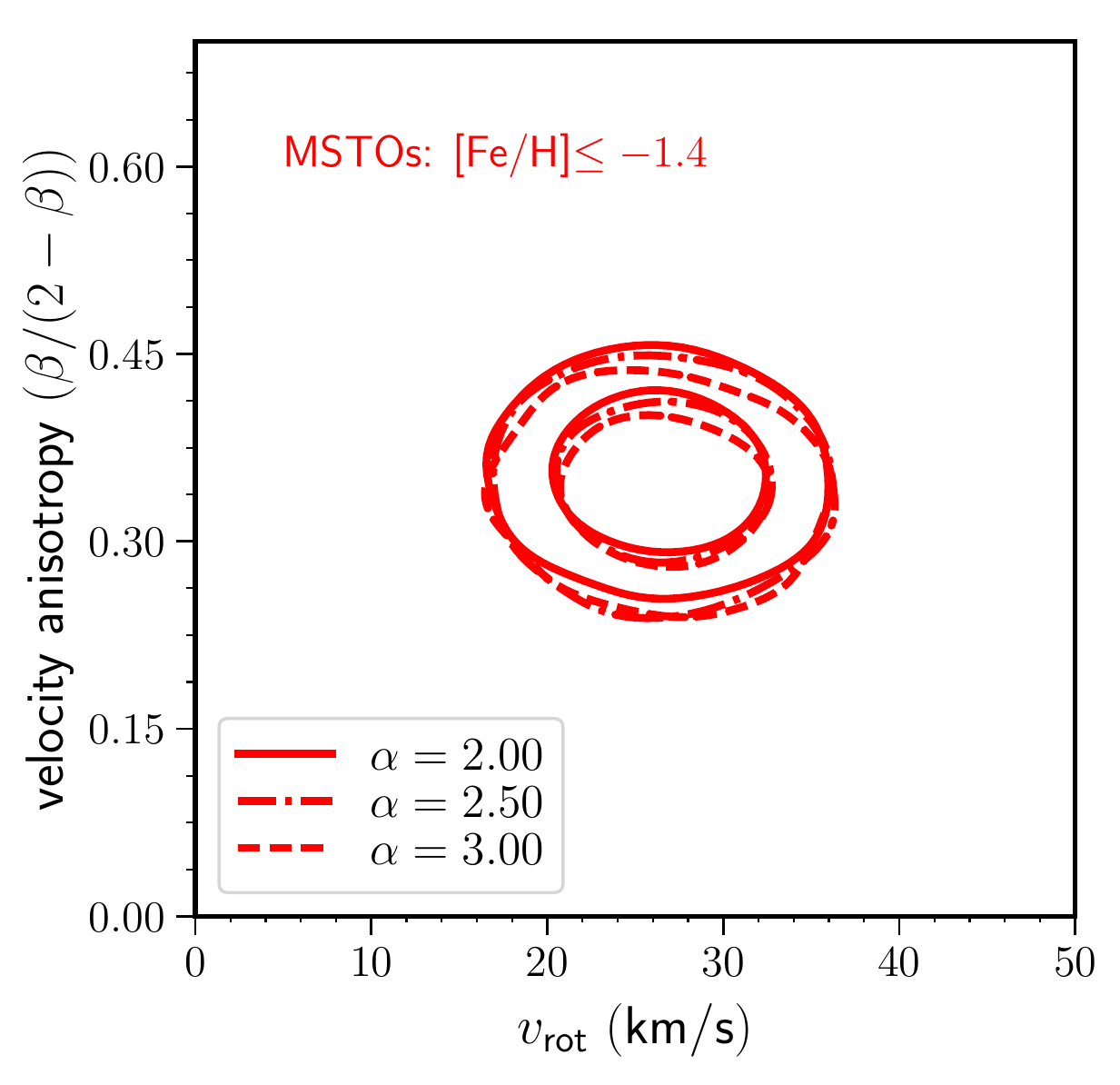}
   \includegraphics[width=0.8\columnwidth]{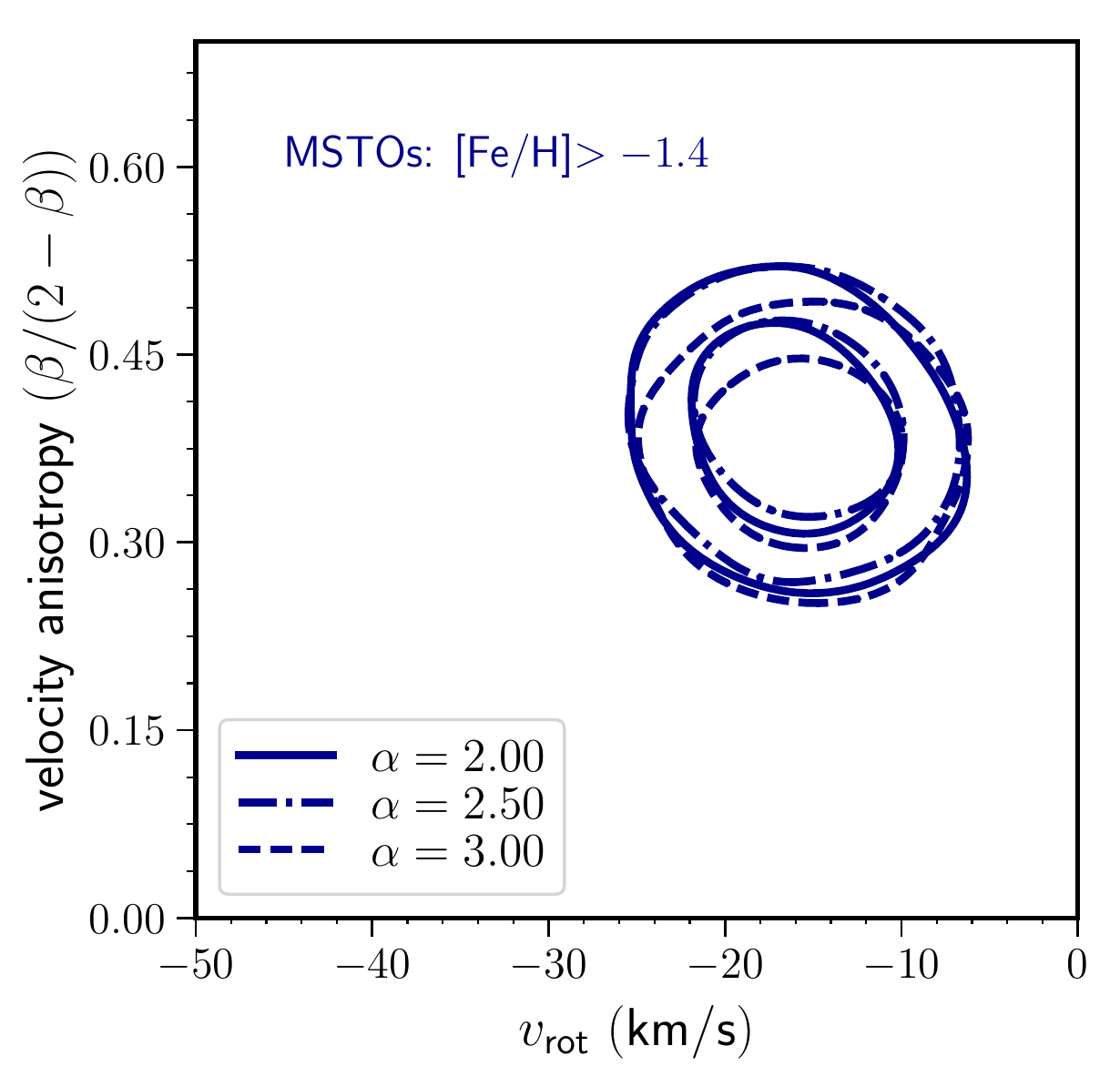}
   \caption{Effect of choice of power-law index $\alpha$ of the radial number density in the measurement of SEGUE MSTOs kinematics.}
   \label{fig:alphaprior}
\end{figure*}
\begin{figure}
   \centering
   \includegraphics[width=1\columnwidth]{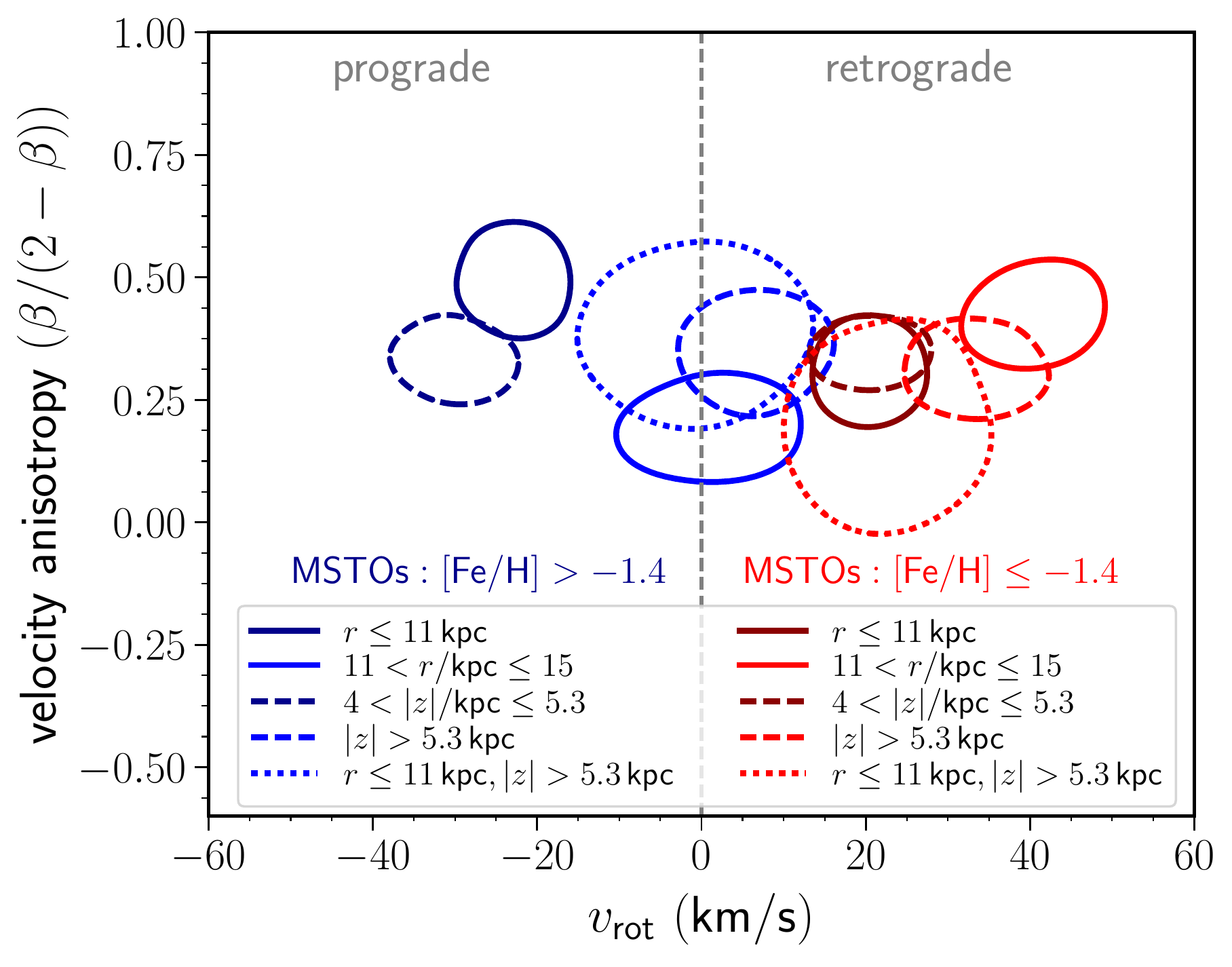}
   \caption{Kinematics of SEGUE MSTOs in radial ($r$) and height above the Galactic plane ($|z|$) bins.
            In cases where limits on r and |z| are not explicitly provided, the limits are $r/\kpc\in[0,15]$ or $|z|/\kpc>4$.}
   \label{fig:rzbin}
\end{figure}
\begin{figure*}
   \centering
   \includegraphics[width=1.7\columnwidth]{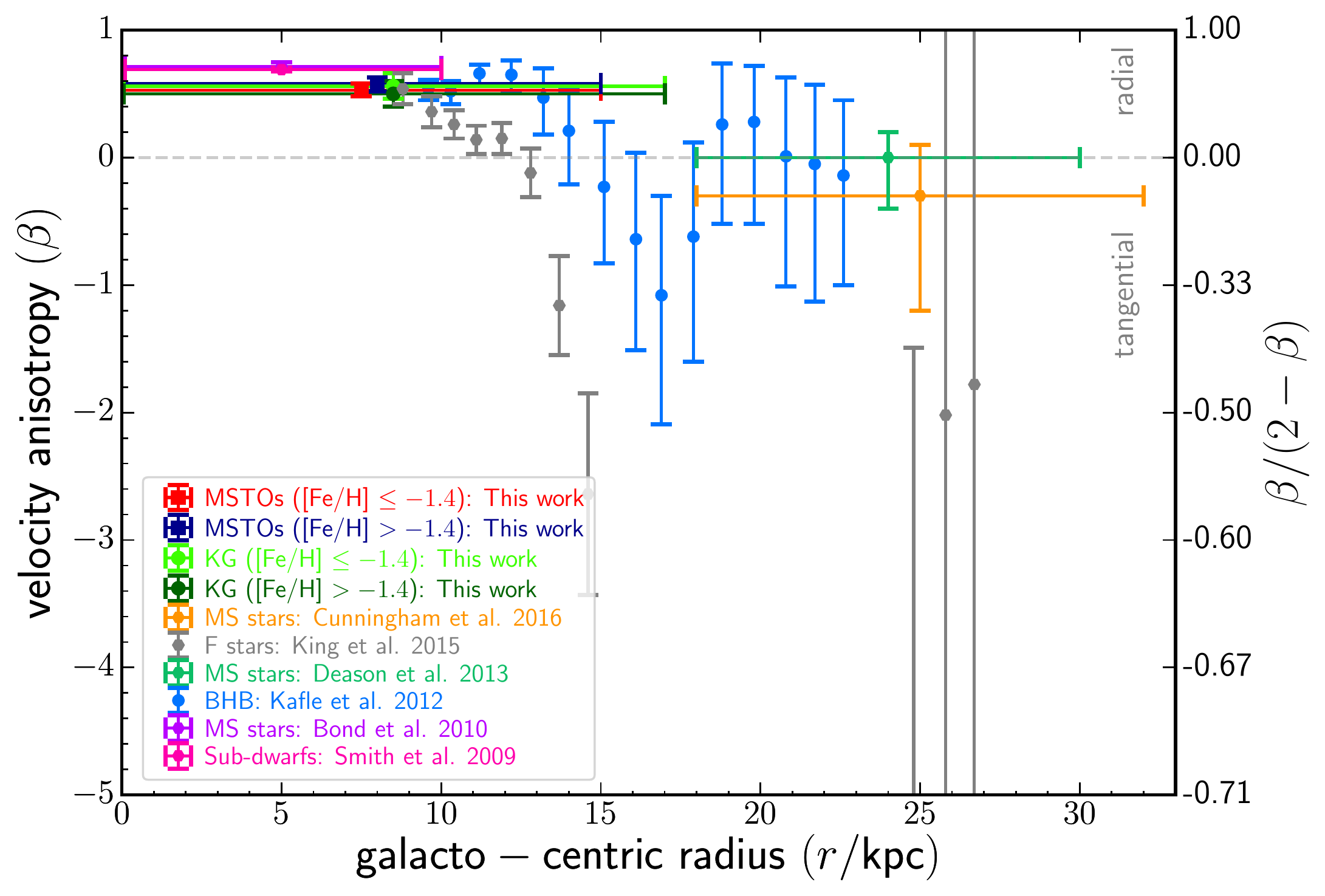}
   \caption{Velocity anisotropy profile of the MW stellar halo for different stellar populations taken from various literature sources as labelled in the figure.
            Red and dark-blue markers show our measurements for the metal-poor and metal-rich MSTOs whereas bright green and dark green markers 
            show our measurements for the metal-poor and metal-rich KGs.}
   \label{fig:betaprofile}
\end{figure*}
\begin{figure*}
   \centering
   \includegraphics[width=1.4\columnwidth]{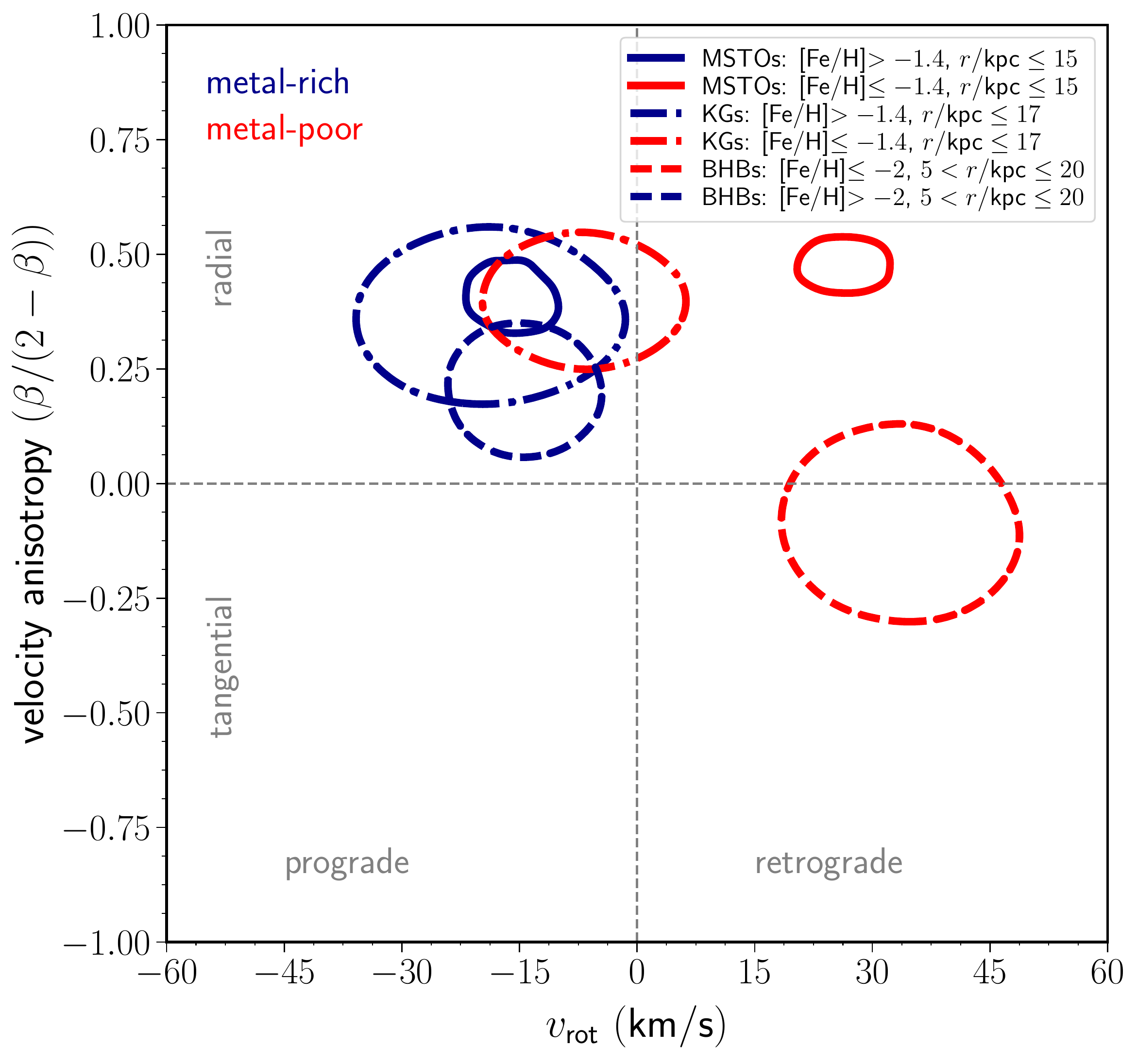}
   \caption{Joint distributions of the modified velocity anisotropy parameter (tangential=-1, radial=1, isotropy=0) and mean azimuthal velocity ($\vrot$) for BHBs, MSTOs and KGs 
            residing the MW stellar halo.}
   \label{fig:betavrot}
\end{figure*}
Now, we apply our scheme to the SEGUE MSTO and KG star catalogues constructed in Section~\ref{sec:data}.
The joint posterior probability distributions of the model parameters ($\rsigma, \thsigma, \phsigma, v_{\rm rot}$) and derived velocity anisotropy ($\beta$) obtained using the MCMC simulation for the MSTOs and KGs are shown in 
Fig.~\ref{fig:mstocorner} and Fig.~\ref{fig:kgcorner} respectively. 
The two red contours show $1\sigma$ and $2\sigma$ credibility intervals for a metal-poor ([Fe/H]$\leq-1.4$) sub-sample.
Likewise the blue contours show the distributions for a comparatively metal-rich ([Fe/H]>-1.4) sub-sample.
As already mentioned previously, we restrict our analysis to $r/\kpc\lesssim15$ for MSTOs and to $r/\kpc\lesssim17$ for KGs.

The most striking feature in the case of MSTOs (Fig.~\ref{fig:mstocorner}) is that there is a clear metallicity dependence in the estimated velocity moments.
We find that the metal-rich and metal-poor samples have opposite mean azimuthal velocity ($v_{\rm rot}$), 
meaning the metal-rich sample shows prograde motion ($-16\pm 4 \kms$) while the metal-poor sample shows retrograde motion ($26\pm4~\kms$). 
The mean difference between their $v_{\rm rot}$ is roughly $40~\kms$.
We also find that the $\rsigma$ and $\thsigma$ of the two sub-samples of MSTOs are quite distinct, whereas the distinction is less clear for $\phsigma$.
Similarly, there is also a small shift in the measured velocity anisotropies of the metal-rich ($\beta=0.58^{+0.06}_{-0.05}$) and metal-poor ($\beta=0.53\pm0.05$) sub-samples, i.e., the MSTOs are clearly radially biased. 
In the case of the KGs (Fig.~\ref{fig:kgcorner}), the distinction is less clear in the $\vrot$ parameter and the velocity dispersions show almost no metallicity bias. 
While the metal-rich KGs do show significant prograde motion ($\vrot=-20\pm{10}~\kms$), 
the metal-poor counterpart is consistent with no-rotation ($\vrot=-7\pm{8}~\kms$) given the uncertainty.
This could be due to the fact that the distances of KGs, which are essentially calibrated with the metallicity dependent colour-absolute magnitude relations obtained from clusters, 
are systematically biased leading to non-rotating metal-poor KGs.
To investigate this we introduce arbitrary systematic shifts of 10 to 25 per cent in the KGs distance and re-run the analysis for the discrepant metal-poor case.
The experiment yields summary kinematics of $\beta=0.55\pm{0.1}$ and $\vrot=-6\pm{9} \kms$.
The estimated $\vrot$ is again consistent with no-rotation.
Recently, \cite{2017arXiv170309230D} also investigated the kinematics of the KG sample obtained from \cite{2014ApJ...784..170X} and combined this with SDSS-\emph{Gaia} proper-motions.
This analysis and our conclusions presented above, both concur that the metal-rich KG ensemble is in pro-grade motion.
Also, though marginal, there is a rotation-metallicity bias in the KGs.

Here we discuss the impact of priors and modelling biases on our final results. 
First, we assess the role of a choice of the power-law slope $\alpha$ of the radial distribution of stars, which enters into our calculation through Equation~\ref{eqn:muprior}.
For this we vary $\alpha$ within a reasonable range $\in[2,3]$; re-analysing all of the above data sets we find that varying $\alpha$ in this range results in negligible changes to the halo kinematics.
An example of this can be seen in Fig.~\ref{fig:alphaprior} where we show 1 and 2 $\sigma$ regions of the joint probability distributions of the summary statistics (modified $\beta$ and $\vrot$) for SEGUE MSTOs for the choices of $\alpha$.
This is because the distance modulus distribution $p(\mu)$ is roughly $\propto s^{3-\alpha}$, which is roughly constant for $\alpha \in [2,3]$.
Similarly, we also re-analysed the metal-rich and metal-poor sub-samples of both the MSTOs and KGs for two additional choices of $R_0/\kpc$: 8.0 and 8.5, which we find has no affect on our kinematic measurements. 
Also, we tested that our final results are robust to the choice of priors on $\rsigma$, $\thsigma$ and $\phsigma$ by $\pm20~\kms$, and $v_l$ and $v_b$  by $\pm100~\kms$.

In this work the model parameters (velocity moments) are assumed to be constant functions of galacto-centric radius ($r$).
This is mainly because we are exploring a small range in the galacto-centric distance $r/\kpc\in[4,15]$.
However, we would also like to investigate if there are any abrupt changes or gradients in the kinematics with distances.
To study this we further split our MSTO catalogue into r, and also $|z|$, bins divided at the median of the distributions ($11~\kpc$ and $5.3~\kpc$ respectively) and analyse each sub-sample separately. 
We could not do the same with KGs due to the small sample size.
Joint probabilities of the summary kinematics (modified $\beta$ and $\vrot$) resulted from this study 
for both the metal-rich and metal-poor MSTO sub-samples are shown in Fig.~\ref{fig:rzbin} and the measurements (velocity moments and also, sample-size) are presented in Table~\ref{table:vmoments}.
We observe that both the metal-rich and metal-poor sub-samples do not show much deviation in velocity anisotropy but the shifts in $\vrot$ are noticeable.
In particular inner sub-samples, i.e., $r/\kpc\leq11$ and $|z|/\kpc\leq5.3$ 
of both the metal-rich (shown in dark-blue solid and dashed lines respectively) and metal-poor (shown in dark red solid and dashed lines respectively)
show leftward shifts, i.e., prograde shifts compare to the respective outer MSTO sub-sample (shown in light blue and light red).
The observed leftward shift could be due to possible disk contaminants.
However, systematic differences in the $\vrot$ among the metal-rich and metal-poor sub-samples remains and retrograde motion of the metal-poor MSTO sub-samples persist.
Albeit less significant, the offset in $\vrot$ is still observed in the case where $r/\kpc\leq11$ and $|z|/\kpc>5.3$ (shown in dotted red and dotted blue).
The large uncertainties here are because of the reduced sample-size.

We provide a list of our final estimates of the model parameters in Table~\ref{table:vmoments}.
Also, as a summary in Fig.~\ref{fig:betaprofile}, we compile the velocity anisotropy of the Galactic halo at different radii taken from various literature sources.
Among the literatures labelled in the figure, 
it is worth noting that \cite{2009MNRAS.399.1223S,2010ApJ...716....1B,2016ApJ...820...18C} utilise the full phase space motion of the stars whereas 
the remaining works \citep{2012ApJ...761...98K, 2013ApJ...766...24D,2015ApJ...813...89K} only use the line-of-sight component of the velocity vector.
The type of stars used in the referred works are also provided in the figure.
The dashed line corresponds to an isotropic velocity distribution, i.e. $\beta=0$. 
It is obviously clear that the $\beta$ profile is not a monotonic function of the galacto-centric radius.
However, it can be stated that the inner stellar-halo of the Galaxy is radial. 
Similarly, to put our results in context, in Fig.~\ref{fig:betavrot} we compare the $\vrot$ and modified velocity anisotropy ($=\beta/(2-\beta)$) estimates for 
metal-rich and metal-poor sub-samples of three different types of stars namely, MSTOs, KGs and BHBs.
The BHB results shown here are obtained from \cite{2013MNRAS.430.2973K}.
Different stellar populations have different metallicity distributions that peak at different [Fe/H], e.g., BHBs are known to be a metal-poor population.
Hence, exact division in the [Fe/H] distribution is not an issue, though note the radial coverage of BHBs is slightly different from that of MSTOs and KGs.
There is a clear segregation in $\vrot$ between metal-poor and metal-rich MSTOs and BHBs, which is less obvious in the case of KGs.
Though BHBs show a clear metallicity bias in anisotropy, it is smaller for MSTOs and negligible for KGs. 
This could be possibly because MSTOs, KGs and BHBs are different stellar populations and might be sampling different accretion events as well as the fractions of these stars sampled by SEGUE varies.

\subsection{Discussion of the systematics}\label{sec:sys}
Finally, we would like to discuss the caveats/limitations of our work that could potentially add systematic biases in our final results.

\emph{Assumption about constant velocity ellipsoid:}
As discussed earlier, a key assumption in our work is that the velocity ellipsoids of the halo stars are aligned along the 
directions of the spherical polar coordinate frame of reference, centred at the Galactic centre. 
While for the inner halo there is evidence supporting the alignment, 
it remains to be seen if the assumption holds beyond the solar-neighbourhood. 
Similarly, because of the dwindling sample size we could only estimate the halo kinematics in coarse radial bins.

\emph{Effects of lack of proper-motions, and potential systematics in $\vlos$:}
Importantly, one of the main limitation of our work is that we only use a line-of-sight ($\vlos$) component of the velocity vector for the stellar halo tracers. 
This is mainly because the variance or the uncertainty of the tangential velocities derived from the proper-motions from the existing astrometric surveys are large.
Moreover, proper motions of stars are known to have significant systematic errors due to frame-dragging and chromatic aberration, which demand a careful treatment \citep{2013MNRAS.432.2402F, 2012MNRAS.420.1281S}.
We note in \cite{2013MNRAS.432.2402F} that even the SEGUE $\vlos$ measurements, in particular for the metal-poor populations, could be systematically offset by $\sim5\,$kms$^{-1}$.
To investigate the effect of the above possible systematic in our final measurements, we add 5\,kms$^{-1}$ in the $\vlos$ for all the metal-poor MSTOs and rerun the MCMC.
We find the introduced systematic shift in the $\vlos$ make no difference in the velocity dispersions of the metal-poor MSTOs kinematics
as we measure $\rsigma=141\pm{3}\,\kms, \thsigma=109\pm{4}\,\kms, \phsigma=82\pm{8}\,\kms$ and $\beta=0.53\pm{0.05}$. 
Moreover, we measure $\vrot=22^{+4}_{-4}\,\kms$, which marginally agree within the $1\sigma$ 
interval with the case of no added systematic (Table~\ref{table:vmoments}).
But the median value of $\vrot$ is found to decrease by $4\,\kms$, which is of the similar order ($5\,\kms$) 
the $\vlos$ has been systematically offset.

\emph{Effects of radial binning:}
In our likelihood analysis, we utilise the full error distribution of the distances.
But occasionally we need to split our data into different spatial, i.e., $r$ and $z$ bins (e.g., in Figure~\ref{fig:rzbin}, or to select halo stars etc) using the point estimates of the distances.
Also, in \cite{2008ApJ...684..287I} we note that a disk-halo transition occurs roughly at $3\,\kpc$, and to select the halo stars we use $|z|>4\,\kpc$ cut.
Nonetheless, due to large distance uncertainties it is possible that the disk-halo separation may not be perfectly cleaned.
As we discuss earlier this could be a reason why we observe that the inner-most halo sample shows relatively 
more prograde motion compared to the sample in the outer bins (dark blue/red contours are left-ward shifted compared to the light blue/red contours in Figure~\ref{fig:rzbin}).
However, even in the outer bins, e.g., a case of $|z|>5.3\,\kpc$, where we assume the halo samples are cleaner, 
the metallicity-kinematics bias persists.

\emph{Effects of potential systematics in \cite{2008ApJ...684..287I} relation:}
We demonstrated in Section~\ref{sec:test} that the $M_r$ calibration shown in Equation~\ref{eqn:Mr} and taken from Ivezic et al. under-predicts $M_r$ for the \galaxia\ MSTOs, on average by +0.37 mag.
While testing the robustness of the Ivezic et al. relation is beyond the scope of this work, in the light of the systematics we observe from \galaxia, it is crucial to investigate its effect on our final results.
For this we utilise the $M_r([\rm{Fe/H}], g_0-i_0)$ relation directly obtained from \galaxia\, instead of the one adopted from Ivezic et al., to calculate distances for SEGUE MSTOs .
The remaining prescription of the analysis are kept exactly same as in the original case presented in Section~\ref{sec:resultsmstokg}.  
In the bottom two rows of Table~\ref{table:vmoments} we provide the kinematics of the SEGUE MSTOs obtained from this exercise.
We observe that both the metal-rich and metal-poor MSTOs sample show increased retrograde motion of $\sim10\,\kms$ in comparison to the original values for $\vrot$.
Note, even with the added systematic in the $M_r$ calibration, we still observe a significant $\vrot$-metallicity bias;
the difference in $\vrot$ of metal-rich and metal-poor sub-samples is $\sim35\,\kms$.
The velocity anisotropies have however decreased from ~0.5 in the original cases to 0.39 for the current samples,
still suggesting mildly radial orbits for the halo stars.

\emph{Effects of (sub)structures present in the Galactic halo:}
It is evident from numerous recent observations that the Galactic halo contains a plethora of coherent stellar debris (substructures) \cite[for review, see][]{2008A&ARv..15..145H,2016ASSL..420...87G}.
Similarly, as we discussed earlier, there are also evidence of the presence of kinematically coherent planar structures of the satellite galaxies in the halo 
\citep{2013Natur.493...62I,2015MNRAS.453.1047P,2015MNRAS.452.1052L,2015ApJ...805...67I}.
It remains to be seen that to what level the un-relaxed stars dispersed from the planar structures or belonging to distinct substructures contribute to the observed kinematic biases in the halo.

\section{Conclusion}\label{sec:conclusion}
In this paper, we model the kinematics of the inner stellar halo ($r\lesssim15~\kpc$) of the Milky Way galaxy
to estimate the mean azimuthal velocity (streaming-motion), velocity dispersion and anisotropy of the main-sequence turn-off (MSTO) and K-giant (KG) stars.
The stellar catalogues are constructed from the Sloan Extension for Galactic Understanding and Exploration (SEGUE) survey \citep{2009AJ....137.4377Y}.
In the following we summarise our main findings:

\begin{itemize}
 \item We find that the comparatively metal-rich sub-samples of all three stellar populations i.e. MSTO (\feh>-1.4), KG (\feh>-1.4) and BHB (\feh>-1.2) 
       inhabiting the MW inner halo are in prograde motion $\sim20\kms$ whereas the metal-poor sub-samples of MSTO and BHB are in a clear retrograde motion of $\sim25\kms$.
       The streaming motion of the metal-poor KGs is also lagging compared to their comparatively metal-rich counterpart.
       However, this lag is significantly smaller than in the cases of the metal-poor MSTO and BHB samples, which are clearly in retrograde orbits.
       
 \item We find that the comparatively metal-rich (\feh>-1.4) and metal-poor (\feh$\leq-1.4$) sub-populations of halo MSTOs show distinct kinematic properties. 
       Importantly, the halo rotation velocity for the sub-populations are offset by $\sim 40 \kms$.
       This is also in agreement with our earlier findings using the BHB stars \citep{2013MNRAS.430.2973K}.
       Some differences in the kinematics of the comparatively metal-rich (\feh>-1.4) and metal-poor (\feh$\leq-1.4$) KGs are also observed but the distinction is not clear-cut.
       Irrespective of the absolute magnitude calibrations obtained from \cite{2008ApJ...684..287I} or derived from \citet[][\galaxia]{2011ApJ...730....3S} to estimate the distances,
       we observe similar level of kinematic-metallicity bias in the MSTOs.
       But, we find that the offset in $\vrot$ is less significant ($~20\,\kms$) when stricter cuts in distances ($r$ and/or $|z|$) are used to cull the disk stars.
       This hints that the net prograde signal is due to disk contamination, particularly the one detected in the metal-rich MSTOs.
       
 \item Both the MSTOs and KGs are on radial orbits (velocity anisotropy $\beta\simeq0.5$), and also velocity anisotropy ($\beta$) does not show any significant metallicity dependence.      
\end{itemize}

In the near future, this work can be extended in two ways.
First, we can utilise the LAMOST survey \citep{2012RAA....12..735D} that provides almost an order-of-magnitude more halo stars of different stellar types e.g. A-type, turn-off, giant etc.
Second, we would immensely benefit from the parallax distances and proper-motions of the inner-halo stars that the second and subsequent data-releases of \emph{Gaia} will deliver. 
Even further down the line, the synergy between \emph{Gaia} and a follow-up radial velocity campaign will produce yet more precise kinematic inference and allow for full phase-space modelling of the Galactic halo.

\section*{ACKNOWLEDGEMENTS}
Funding credit:
PRK is funded through Australian Research Council (ARC) grant DP140100395 and The University of Western Australia Research Collaboration Award PG12104401 and PG12105203.

Software credit: 
\href{https://doi.org/10.5281/zenodo.225727}{GetDist} \citep{2002PhRvD..66j3511L},
{\sc ipython} \citep{ipython}, {\sc matplotlib} \citep{matplotlib}, {\sc numpy} \citep{numpy},
{\sc pandas} \citep{pandas} and {\sc scipy} \citep{scipy}.

Content credit: 
We like to thank the referees, and Pascal Elahi for their useful comments.

Funding for SDSS-III has been provided by the Alfred P. Sloan Foundation, the Participating Institutions, the National Science Foundation, 
and the U.S. Department of Energy Office of Science. The SDSS-III web site is http://www.sdss3.org/.

\appendix

\section{Transformation of phase-space coordinates}\label{sec:coordtransform}
Here we provide the transformation formulary to convert phase-space coordinates from the galactic (centred at the Sun) 
to the galacto-centric (centred at the centre of the Galaxy) frame of reference, i.e.
\begin{equation}
(l,b,d_{\rm gc}, v_l, v_b, \vlos) \longrightarrow (r_{\rm GC}, \theta_{\rm GC}, \phi_{\rm GC}, v_r, v_\theta, v_\phi). 
\end{equation}

First, we assume a generic basis vector transformation matrix
\begin{equation}
\mathbf{T}(\theta, \phi)=
  \begin{bmatrix}
    \sin(\theta) \cos(\phi) & \sin(\theta)\sin(\phi) & \cos(\theta)\\
    \cos(\theta) \cos(\phi) & \cos(\theta)\sin(\phi) & -\sin(\theta)\\
    -\sin(\phi) & \cos(\phi) & 0    
  \end{bmatrix}
\end{equation}

To transform from spherical galactic coordinates ($l,b,d_{\rm gc}$) to the cartesian coordinates we use
\begin{equation}
  \begin{bmatrix}
    x_{\rm gc}\\
    y_{\rm gc}\\
    z_{\rm gc}    
  \end{bmatrix}
= \mathbf{T^{-1}}(90^\circ-b, l)
  \begin{bmatrix}
    0\\
    0\\
    d_{\rm gc},
  \end{bmatrix}
\end{equation}
where, $d_{\rm gc}$ is a helio-centric distance to a star.
Similarly, velocities in spherical galactic coordinates ($v_l,v_b,\vlos$) can be converted to cartesian coordinates as following
\begin{equation}
  \begin{bmatrix}
    vx_{\rm gc}\\
    vy_{\rm gc}\\
    vz_{\rm gc}    
  \end{bmatrix}
= \mathbf{T^{-1}}(90^\circ-b, l)
  \begin{bmatrix}
    \vlos\\
    -v_b\\
    v_l.
  \end{bmatrix}
\end{equation}
Now, we linearly shift the galactic phase-space cartesian coordinates to the centre of the Galaxy using
\begin{equation}
\begin{split}
\{ x_{\rm GC}, y_{\rm GC}, z_{\rm GC} \} = \{ & x_{\rm gc} - R_0, y_{\rm gc}, z_{\rm gc} \} \\
\{ vx_{\rm GC}, vy_{\rm GC}, vz_{\rm GC} \} = \{ & vx_{\rm gc} + U_\odot, vy_{\rm gc}+ V_\odot + v_{\rm LSR},\\
& vz_{\rm gc}+ W_\odot \},
\end{split}
\end{equation}
where we are ignoring the height of the Sun from the plane of the galactic disk, which would otherwise introduce an additional term in $z_{\rm gc}$.

Finally, we can convert the galacto-centric cartesian to spherical coordinates as followings
\begin{equation}
 \begin{split}
 r_{\rm GC} & = \sqrt{x_{\rm GC}^2 + y_{\rm GC}^2 +z_{\rm GC}^2} \\
 \theta_{\rm GC} & = \cos^{-1} z_{\rm GC}/r_{\rm GC}\\
 \phi_{\rm GC} & = \tan^{-1} y_{\rm GC}/x_{\rm GC}
 \end{split}
\end{equation}
and 
\begin{equation}
  \begin{bmatrix}
    v_r\\
    v_\theta\\
    v_\phi    
  \end{bmatrix}
= \mathbf{T}(\theta_{\rm GC}, \phi_{\rm GC})
  \begin{bmatrix}
    vx_{\rm GC}\\
    vy_{\rm GC}\\
    vz_{\rm GC}
  \end{bmatrix}
\end{equation}
In case the heliocentric equatorial coordinates and/or proper-motions are to be used, one could refer to 
\cite{1987AJ.....93..864J,2005MNRAS.359.1287B} for the extra steps that need to be undertaken.

\bibliographystyle{mnras}
\bibliography{paper.bbl}

\bsp	\label{lastpage}
\end{document}